\documentclass[prb]{revtex4}

\usepackage[T1]{fontenc}

\usepackage[thinspace,squaren,pstricks,italian,derivedinbase,derived]{SIunits}
\addunit{\bohr}{Bh}
\addunit{\rydberg}{Ry}
\addunit{\chwilka}{ch}
\addunit{\kubo}{Kb}

\usepackage[latin2]{inputenc}
\usepackage[dvips]{epsfig}
\usepackage{amsfonts}
\usepackage{amsmath}
\usepackage{amssymb}
\usepackage{upgreek}  
\usepackage{mathrsfs}
\usepackage{booktabs} 
\usepackage{csquotes}

\renewcommand{\vec}[1]
{
{\mathbf #1}
}
\newcommand{\abs}[1]
{
{\left| #1 \right|} 
}

\newcommand{\Ry}
{
{Ry}
}

\newcommand{\eps}
{
{\epsilon}
}

\newcommand{\onemat}
{
{\sf I}     
}

\newcommand{\Heav}
{
{\theta}
}

\renewcommand{\Im}
{
{\mathfrak{Im}}
}
\newcommand{\ii}
{
{{\mathfrak{i}}} 
}

\newcommand{\av}[1]
{
{\left\langle #1 \right\rangle} 
}
\newcommand{\sqbr}[1]
{
{\left[ #1 \right]}
}

\newcommand{\fbr}[1]
{
{\left( #1 \right)} 
}

\newcommand{\suscSymb}
{
\chi
}

\newcommand{\comm}[3]
{
{\sqbr{#1,#2}_{#3}}
}

\newcommand{\Psigma}
{
{\boldsymbol{\sigma}}
}

\newcommand{\fFD}
{
{f_{T}}
}

\newcommand{\zp}
{
{0^{+}}
}

\newcommand{\xc}
{
_{\mathrm{xc}}
}
\newcommand{\KS}
{
_{\mathrm{KS}}
}

\newcommand{\el}
{%
\textit{et~al}.\
}

\newcommand{\Angstrom}{\textrm{\AA}}

\newcommand{\Loss}{\mathcal{L}}

\newcommand{\vC}{v_{\textrm{C}}}

\begin{document}


\title{First-principles perspective on magnetic second sound}
\author{Pawe\l{} Buczek}
\email[Corresponding author: ]{pawel.buczek@haw-hamburg.de}
\affiliation{Department of Engineering and Computer Sciences, \\Hamburg University of Applied Sciences, Berliner Tor 7, 20099 Hamburg, Germany}
\author{Nadine Buczek}
\affiliation{Department of Applied Natural Sciences, L\"ubeck University of Applied Sciences, M\"onkhofer Weg 239, 23562 L\"ubeck, Germany}
\author{Giovanni Vignale}
\affiliation{Department of Physics and Astronomy, University of Missouri, 223 Physics Building, Columbia, Missouri 65211, USA}
\author{Arthur Ernst}
\affiliation{Institute for Theoretical Physics, Johannes Kepler University Linz, Altenberger Stra\ss{}e 69, 4040 Linz, Austria}
\affiliation{Max-Planck-Institut of Microstructure Physics, Weinberg 2, 06120 Halle (Saale), Germany}

\date{\today}

\begin{abstract}
The fluctuations of the magnetic order parameter, or longitudinal spin excitations, are investigated theoretically in the ferromagnetic Fe and Ni as well as in the antiferromagnetic phase of the pnictide superconductor FeSe. The charge and spin dynamics of these systems is described by evaluating the generalized charge and spin density response function calculated from first-principles linear response time dependent density functional theory within adiabatic local spin density approximation.
We observe that the formally non-interacting Kohn-Sham system features strong coupling between the magnetization and charge dynamics in the longitudinal channel and that the coupling is effectively removed upon the inclusion of the Coulomb interaction in the charge channel and the resulting appearance of plasmons. The longitudinal spin fluctuations acquire a collective character without the emergence of the Goldstone boson, similar to the case of paramagnon excitations in non-magnetic metals like Pd. In ferromagnetic Fe and Ni the longitudinal spin dynamics is governed by interactions between low-energy intraband electron-hole pairs while in quasi two dimensional antiferromagnet FeSe it is dominated by the interband transitions with energies of the order of exchange splitting. In the later material, the collective longitudinal magnetization fluctuations feature well defined energies and long life times for small momenta and appear below the particle-hole continuum. The modes become strongly Landau-damped for growing wave-vectors. We relate our theoretical findings to existing experimental spin-polarized electron energy loss spectroscopy results. In bulk bcc Fe, the longitudinal magnetic modes appear above the typical energies of transverse spin-waves, have energies comparable with the Stoner spin-flip excitation continuum, and are order of magnitude less energetic than the charge dynamics.

\end{abstract}

\maketitle

\section{Introduction}
\label{sec:Introduction}

The spin fluctuations in itinerant (metallic) system attract unquenched interest of theoretical, experimental, and engineering communities as their properties remain central to the full understanding of a broad spectrum of physical and technical problems. The magnetic excitations drive the phase transitions at Curie and N\'eel temperatures \cite{Pajda2001,Rusz2005} as well as in the vicinity of quantum critical points \cite{Brando2016}. They give rise to unconventional coupling mechanisms in new families of high temperature superconductors \cite{Mazin2008,Mazin2009,Essenberger2014,Linscheid2015a,Essenberger2016} and form the basis for information processing in spintronic computers \cite{Zutic2004,Kajiwara2010,Khitun2010,Joshi2016}. Owing to the recent impressive experimental progress they can be probed even in the nanostructures \cite{Zakeri2013,Balashov2014,Zakeri2014,Qin2015}.

Despite the enormous relevance, is it surprising to observe that an important and intriguing class of spin density excitations, the so called \textit{longitudinal spin fluctuations}, received little attention in the literature so far. These excited states constitute the central theme of this paper. As they are somewhat exotic and less known member of the spin excitations family, we will briefly put them in the context of the theory of itinerant magnets before proceeding with their detailed microscopic analysis.

In a simple picture, the itinerant magnets undergo a phase transition at respective critical temperature associated with the emergence of long range magnetic order (the order parameter being macroscopic magnetization) and spontaneous rotational and time reversal symmetry breaking \cite{Anderson1990,Peierls1991,Tscheuschner1992}. The broken symmetry results in the appearance of a Goldstone boson and the corresponding family of low energy excitations, the spin waves or magnons. These excited states involve a spin flip and are essentially associated with a tilt, i.e.\ a change of the \textit{direction} of the local magnetization with respect to the ground state direction. (We consider only collinear magnets in this work and assume the magnetization $m^{z}$ to point along the $z$-axis.) The magnons in collinear magnets -- in the absence of spin orbit interaction -- involve solely fluctuations of the magnetization in the direction perpendicular to the $z$-axis which are practically decoupled from the charge excitations.

At temperature $T = 0$, the longitudinal spin excitations are associated with the variation of \textit{value} of the longitudinal ($z$-component) magnetization density. The symmetry of the collinear band structure, i.e. the one involving up and down bands of different dispersion, requires that such fluctuations couple to the charge density fluctuations, including plasmons. (We discuss this issue in more details in Sec.~\ref{sec:Formalism}). Little is known about the exact physical picture of the coupling. Furthermore, questions concerning the existence of collective modes in this channel and, if they indeed exist, their dispersion and life-times have hardly been addressed, in particular for complex systems like antiferromagnets.

On the other hand, the longitudinal spin excitations of itinerant magnets relate to one of the most exciting effects in the many body physics, namely the emergence of the ``second sound'' in the superfluids, in particular in the liquid helium \cite{Tisza1938, Landau1941, Dingle1950}. Since in a magnet the expectation value of the $z$-component of the magnetization plays the role of the order parameter, the longitudinal spin excitations are in fact the fluctuations of the order parameter themselves. Similarly, the second sound in liquid helium involves the fluctuations of the boson condensate density being the superfluid phase order parameter.  However, despite this apparent relation, the two manifestations of equivalent phenomenon exhibit rather different underlying microscopic pictures and we defer the detailed discussion of the analogy to the summary in Subsec.~\ref{sec:Summary}.

The second sound in helium is associated with the heat transport which, unlike typically in usual matter, occurs by wave propagation rather than by diffusion. Apart from the liquid helium, the second sound has been observed in ultracold atomic gases \cite{Sidorenkov2013}. Such wave-like heat transport emerges is solids as well \cite{Pitaevskii1968, Hardy1970, Huberman2019}. As pointed out by Chester \cite{Chester1963} this phenomenon appears when the heat flow is unable to infinitely quickly respond to the temperature gradient, leading to the appearance of the time derivative of the thermal heat current in a generalized Fourier equation, thus giving its solution the wave-like character. Equivalent effects arise in the cold Fermi gases where ballistic (wave-like) and diffusive transport can coexist, e.g. correspondingly for charge and spin excitations \cite{Polini2007, Xianlong2008}. It must be noted, however, that the second sound in solids involves no fluctuations of any condensate density or order parameter.

However, the time resolved dynamics of order parameters can be accessed in several sophisticated experiments. The fluctuating Cooper pair density, the latter being the order parameter in the superconductors, has been studied \cite{Littlewood1982, Goldman2006, Barlas2013, Morice2018}. Signatures of the longitudinal spin fluctuations can be experimentally observed in thermal neutron scattering \cite{Koetzler1986, Boeni1991, Schweika2002, Wang2016, Inosov2016, Ma2017, Chen2019} and in the spin-polarized electron energy loss spectroscopy experiments (SPEELS) \cite{Kirschner1985, Venus1988, Vasilyev2016}.

Concerning the outlined broad many-body context of the longitudinal spin fluctuations, it is exciting to explore whether well defined collective excitations occur in this channel and, in case they do, what are their properties. First principles theoretical investigations of the longitudinal spin excitations in itinerant magnets are scarce and this state of affairs is largely attributable to the lack of suitable theoretical tools. The bulk of investigations of the transverse spin excitations (spin-waves) were performed in the framework of the adiabatic mapping onto the Heisenberg Hamiltonian \cite{Liechtenstein1987, Sandratskii1991, Sandratskii1998, Halilov1998a, Antropov1999, Grotheer2001, Buczek2016, Buczek2018}. In this approach, at absolute zero, there are strictly no longitudinal magnetic excitations, as the moments in the Heisenberg model are treated as rigid entities. The spectrum of spin fluctuations -- both transverse as well as longitudinal -- can be inferred form the frequency and wave-vector dependent magnetic susceptibility $\chi(\vec{q},\omega)$ which itself is amenable to the \textit{ab initio} treatment by means of the linear response time dependent density functional theory \cite{Gross1985}. Unfortunately, the corresponding necessary numerical calculations are characterized by high algorithmic complexity and non-negligible computational costs \cite{Sasiouglu2010, Rousseau2012, SantosDias2015, Cao2018, Gorni2018, TancogneDejean2020}. Only recently, Wysocki \el \cite{Wysocki2017} evaluated the spectrum of longitudinal spin excitations for elementary $3d$ transition metals. In course of recent years, we have developed a highly efficient numerical scheme for the computation of the linear response functions of solids within the time dependent density functional theory based on the Korringa-Kohn-Rostoker (KKR) Green's function (GF) method \cite{Buczek2011a} and applied it to complex bulk materials \cite{Buczek2009} and thin films \cite{Schmidt2010a, Buczek2010d, Buczek2011, Zakeri2013, Qin2015}. The scheme is general and applicable to all collinear magnets, including ferro- and ferrimagents \cite{Sandratskii2012, Odashima2013} as well as paramagnetic phases \cite{Essenberger2012}. Here, we report on the extension of the scheme to the examination of longitudinal and charge excitations.

In this paper, we first discuss the longitudinal spin excitations in Fe and Ni. We aim to provide additional insights into the microscopic nature of the fluctuations in these systems and open the possibility to contrast the spin dynamics in ferromagnets with the one of antiferromagnets. Thus, we proceed to the more involved case of the iron-based superconductor FeSe. There exists a compelling evidence that in this system the superconductivity is mediated by the spin excitations of the non-magnetic phase (so called paramagnons) \cite{Lischner2015, Essenberger2016}. Unfortunately, the local spin density approximation (LSDA) predicts a stable antiferromagnetic ground state phase which contradicts the experiment \cite{Sharma2018}. It is generally believed that the deficiency of the LSDA stems from its failure to capture the detrimental influence of the intense spin fluctuations induced by a hidden quantum critical point on the magnetic ordering in such systems \cite{Wysocki2011, Essenberger2012}. The spectrum of the spin fluctuations in the ordered phase is thus the key ingredient for an improvement of the first principle predictions of the ground state. It seems to be a general observation for many itinerant systems also beyond the pnictide family of superconductors. For example, as shown by Derlet \cite{Derlet2012}, the inclusion of the longitudinal spin fluctuations is unexpendable in the realistic modeling of magnets with strongly polarizable constituents.

The paper is organized as follows. In Sec.\ \ref{sec:Formalism} the formalism of the time dependent density functional theory allowing for the calculations of the charge and longitudinal spin susceptibility is exposed. The results of the numerical studies are presented and discussed in Sec.\ \ref{sec:Results}.

\section{Formalism}
\label{sec:Formalism}

Our aim is to describe the spectrum of its excited states involving longitudinal fluctuations of the magnetization. In order to achieve it, we resort to the evaluation of the retarded generalized charge and magnetization susceptibility \cite{Buczek2011a}
\begin{align}
  \chi^{ij} \fbr{\vec{x},\vec{x}',t-t'} &= - \ii \Heav(t-t')
    \av{\comm%
      {\hat{\sigma}^{i}\fbr{\vec{x}t}}
      {\hat{\sigma}^{j}\fbr{\vec{x}'t'}}
    {}},
\label{eq:GeneralDensityResponse}
\end{align}
which relates the linear charge or magnetization density response $\delta n^{i}\fbr{x}$ of the system under consideration to the applied dynamical magnetic or scalar field $\delta V^{j}\fbr{\vec{x}'}$. $\hat{\sigma}^{i}$ are charge ($i = 0$) and magnetization density operators ($i = x, y, z$), $\comm{A}{B}{}\equiv AB - BA$, and the $\av{\hat{o}}$ is the expectation value of the operator $\hat{o}$ for the unperturbed system.

The time Fourier transformation of the susceptibility, $\chi^{ij} \fbr{\vec{x},\vec{x}',\omega}$, has a clear physical interpretation following the \emph{fluctuation dissipation theorem} \cite{Nyquist1928,Callen1951,Kubo1966}. Its non-zero imaginary part for certain frequency $\omega$ signifies the presence of excited states of the underlying unperturbed Hamiltonian with this energy\footnote{Unless otherwise specified, Rydberg
atomic units are used throughout, with $\hbar = 1$, Bohr radius $a_{0} = 1$, and Rydberg energy $E_{\mathrm{R}} = 1$ which in turn implies the numerical values of the electron charge and mass to be respectively $e = \sqrt{2}$ and $m_{e} = \frac{1}{2}$.} and involving fluctuating charge and magnetization densities. For complex solids, considering all density channels at once, the imaginary part must be generalized as \emph{loss matrix}, i.e.\ the anti-Hermitian part of the susceptibility
\begin{align}
 \Loss\sqbr{\chi^{ij}(\vec{x},\vec{x}',\omega)} \equiv \frac{1}{2\ii}
    \fbr{\chi^{ij}(\vec{x},\vec{x}',\omega) - \chi^{ji}(\vec{x}',\vec{x},\omega)^{\ast}}.
\end{align}
For a chosen energy $\omega$, the eigenvectors $\xi_{\lambda}(\vec{x})$ of $\Loss\sqbr{\chi^{ij}}$ represent the \emph{shapes} of the resonant \emph{natural modes} (charge and magnetization density fluctuations) at this frequency. The magnitude of the associated eigenvalues, $\Loss\sqbr{\chi^{ij}}_{\lambda}$, give the intensity (density of states) of these modes.

The susceptibility can be obtained from the linear response time dependent density function theory (LRTDDFT) \cite{Gross1985,Qian2002} in a two step procedure. First, the \textit{Kohn-Sham susceptibility}
\begin{align}
\suscSymb^{ij}\KS\fbr{\vec{x},\vec{x}',\omega} = 
  \sum_{km} \sigma^{i}_{\alpha\beta} \sigma^{j}_{\gamma\delta} 
    \fbr{f_{k} - f_{m}}
    \frac{
      \phi_{k}\fbr{\vec{x}\alpha}^{\ast}
      \phi_{m}\fbr{\vec{x}\beta}
      \phi_{m}\fbr{\vec{x}'\gamma}^{\ast}
      \phi_{k}\fbr{\vec{x}'\delta}
  }
  {
    \omega + \fbr{\eps_{k} - \eps_{m}} + \ii 0^{+}
  },
\label{eq:KSSusc}
\end{align}
yields the density response, or equivalently single-particle excitation spectrum, of the formally non-interacting Kohn-Sham (KS) system. In the above equation $\phi_{j}\fbr{\vec{x}\alpha}$'s and $\eps_{j}$'s denote respectively KS eigenfunctions and corresponding eigenenergies. $f_{j} \equiv \fFD(\eps_{j})$, where $\fFD(\eps)$ is the Fermi-Dirac distribution function. Second, when an external field is applied,the induced charge and magnetization densities described by the $\suscSymb^{ij}\KS$ alter the Hartree and exchange-correlation potential leading to a self-consistent \emph{susceptibility Dyson equation}
\begin{align}
  \chi^{ij} \fbr{\vec{x},\vec{x}',\omega} = \chi^{ij}\KS \fbr{\vec{x},\vec{x}',\omega} +
    \sum_{k,l=0}^{3}
    \iint d\vec{x}_{1}d\vec{x}_{2} \chi^{ik}\KS \fbr{\vec{x},\vec{x}_{1},\omega}
      \fbr{
        K\xc^{kl}\fbr{\vec{x}_{1},\vec{x}_{2},\omega}
        + \frac{2 \delta_{k0}\delta_{l0}}{\abs{\vec{x}_{1} - \vec{x}_{2}}}
      }
      \chi^{lj}\fbr{\vec{x}_{2},\vec{x}',\omega}
\label{eq:GeneralDyson}
\end{align}
which allows to find the true interacting (enhanced) susceptibility of the many-body system providing that the exchange-correlation kernel, $K\xc$, defined as a functional derivative of exchange-correlation potential evaluated at the ground state values of electronic and magnetic densities
\begin{align}
  K\xc^{ij}\sqbr{\av{\hat{\Psigma}\fbr{\vec{x}}}}\fbr{\vec{x},\vec{x}',t-t'} \equiv
    \frac{\delta v\xc^{i}\fbr{\vec{x},t}}{\delta n^{j}\fbr{\vec{x}'t'}},
\end{align}
is known. Furthermore, in what follows, we denote the Hartree (Coulomb) interaction with $\vC\fbr{\vec{x}} = 2/x$.

The determination of the exchange-correlation kernel is equivalent to the exact solution of the many-body problem and equally difficult. In this work we resort to the adiabatic local spin density approximation (ALSDA)
\begin{align}
  K\xc^{ij}\sqbr{\av{\hat{\Psigma}\fbr{\vec{x}}}}\fbr{\vec{x},\vec{x}',t - t'} \approx
    \frac{\delta v_{\textrm{LSDA}}^{i}\sqbr{\av{\hat{\Psigma}\fbr{\vec{x}}},\vec{x}}}{\delta n^{j}\fbr{\vec{x}}}
    \delta\fbr{\vec{x} - \vec{x}'} \delta\fbr{t - t'}.
\end{align}
Recently, there has been a progress in constructing non-local magnetic exchange-correlation functionals \cite{Kurth2009,Eich2010,Eich2013,Eich2013a,Pittalis2017,Eich2018}, but their their inclusion in practical LRTDDFT calculations is still an ongoing effort.

Let us consider first the structure of the Kohn-Sham susceptibility. Within the non-relativistic local spin density approximation (LSDA), the Kohn-Sham states of the collinear ferro-, antiferro-, and paramagnetic systems can be characterized by a certain value of the spin-projection. We adopt the convention that the ground-state magnetization $\vec{m}\fbr{\vec{x}}$ points everywhere along the $\pm z$ direction which we select as the axis of the spin-quantization. In this case the susceptibility $\chi^{ij}\KS$ has only four independent elements and the following structure
\begin{align}
\chi\KS =
  \begin{pmatrix}
     \phantom{-}\suscSymb\KS^{xx} & \suscSymb\KS^{xy} &                 0  &                0 \\
              - \suscSymb\KS^{xy} & \suscSymb\KS^{xx} &                 0  &                0 \\
                               0 &                0 &  \suscSymb\KS^{00}  & \suscSymb\KS^{0z} \\
                               0 &                0 &  \suscSymb\KS^{0z}  & \suscSymb\KS^{00}
  \end{pmatrix}
\end{align}
in which the transverse (ie.\ with the direction in the $xy$-plane) magnetic fluctuations are strictly decoupled from the fluctuations of the $z$ component of magnetization and the charge fluctuations. The transverse susceptibilities, $\suscSymb^{\pm} = \suscSymb^{xx} \mp \ii \suscSymb^{xy}$, describe the spin-flip processes, respectively up-to-down ($+$) and down-to-up ($-$). We refer to the as Stoner and anti-Stoner excitations. On the other hand, the longitudinal susceptibilities ($i,j=0,z$) involve particle-hole pairs of the same spin. The longitudinal block can be readily diagonalized and the eigenvalues of the susceptibility read
\begin{align}
	\suscSymb\KS^{00} &= \suscSymb\KS^{zz} = \suscSymb\KS^{\uparrow} + \suscSymb\KS^{\downarrow}, \\
	\suscSymb\KS^{0z} &= \suscSymb\KS^{z0} = \suscSymb\KS^{\uparrow} - \suscSymb\KS^{\downarrow},
\end{align}
where $\suscSymb\KS^{\uparrow,\downarrow}$ are the response functions of formally non-interacting spin up and down bands. Thus, it turns out, that the $m_{z}$-charge coupling is given by the \emph{difference} in the dynamics of the up and down electron channels. As such, it vanishes for paramagnets where the bands are degenerate but, interestingly, not for antiferromagnets. Despite the fact that in an antiferromagnet the up and down bands are degenerate as well, $\suscSymb\KS^{z0}$ does not vanish due to the fact that these two bands have different spatial characters.

We observe that the decoupling between the transverse and longitudinal fluctuations is preserved in the enhanced susceptibility structure when the $K\xc$ is approximated within the ALSDA. Furthermore, due to the presence of the Coulomb term in the charge channel of the susceptibility Dyson equation \eqref{eq:GeneralDyson}, the charge and longitudinal magnetization responses cease to become identical ($\suscSymb^{00} \ne \suscSymb^{zz}$), also for paramagnets.

Concerning our computational scheme, it is exposed in a great detail in Ref.~\cite{Buczek2011a}. Both the KKR Green's function and the susceptibilities are expanded in real spherical harmonics up to $l_{\mathrm{max}} = 3$. We use about 100 energy points in the complex energy integration contour. For the convolution of two Green's functions, the adaptive sampling of the Brillouin zone is used which uses between 64 (away from the real axis) and $2\times 10^{5}$ (close to the real axis) $k$-points. 12 Chebyshev polynomials are used to describe the dependence of the susceptibilities in the atomic cells. All quantities are carefully converged with respect to these parameters.

\section{Results}
\label{sec:Results}

\subsection{Elementary ferromagnets Fe and Ni}
\label{subsec:ElementaryFerromagnetsFeAndNi}

\begin{figure}[htbp]
	\centering
		\includegraphics[width=0.50\textwidth]{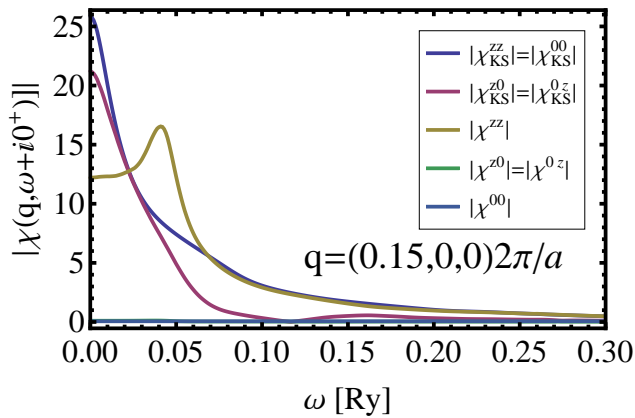}
	\caption{Magnitudes of the response functions in the longitudinal channel in Ni for a selected momentum transfer. The picture does not qualitatively change for other momenta and is similar to the Fe case as well. For simplicity, in the response function, it is assumed that the driving field is uniform in the atomic cell of the atom and the response is integrated over the cell. The terms $\suscSymb^{0z} = \suscSymb^{z0}$ and $\suscSymb^{00}$ are practically zero in the scale of the figure.}
	\label{fig:susc-Ni-abs}
\end{figure}

In this section we study the longitudinal spin fluctuations in bcc Fe and fcc Ni. Let us first outline the general picture of the electron dynamics in the longitudinal channel. As discussed in Section~\ref{sec:Formalism}, the formally non-interacting Kohn-Sham electron features two distinct normal modes $\suscSymb\KS^{\uparrow}$ and $\suscSymb\KS^{\downarrow}$ associated with the decoupled dynamics of formally non-interacting up and down electron channels. In principle, both of these modes involve coupled charge and magnetization dynamics and the strength of the coupling is given by the magnitude of the $\suscSymb\KS^{0z}$ susceptibility. As evident from Figure~\ref{fig:susc-Ni-abs}, the magnetization-charge interaction is indeed substantial as its magnitude compares to the one of the same-channel $\suscSymb\KS^{00} = \suscSymb\KS^{zz}$ response functions. 

The pictures changes qualitatively when the true (enhanced) susceptibility is obtained from the susceptibility Dyson equation. The Coulomb interaction active in the $\suscSymb^{00}$ channel causes the charge (plasmon) dynamics to develop at characteristic energy scale of several \Ry \cite{Gurtubay2005}. As a result, the low energy window, cf.\ Figure~\ref{fig:susc-Ni-abs}, is clearly dominated by the longitudinal spin dynamics, given by $\suscSymb^{zz}$, with practically vanishing coupling to the charge density excitations, $\suscSymb^{0z} = \suscSymb^{z0} 	\approx 0$. Owing to this separation of the energy scales, the longitudinal spin dynamics in ferromagnets becomes qualitatively similar to the one in the transverse channel, given by $\suscSymb^{\pm}$, which also does not involve the coupling to the charge excitations. We stress, however, that in the ALSDA without spin-orbit interaction, the transverse-longitudinal and the transverse-charge decoupling is exact whereas the charge-longitudinal decoupling is only approximate.

\begin{figure}[htbp]
	\centering
		\includegraphics[width=0.50\textwidth]{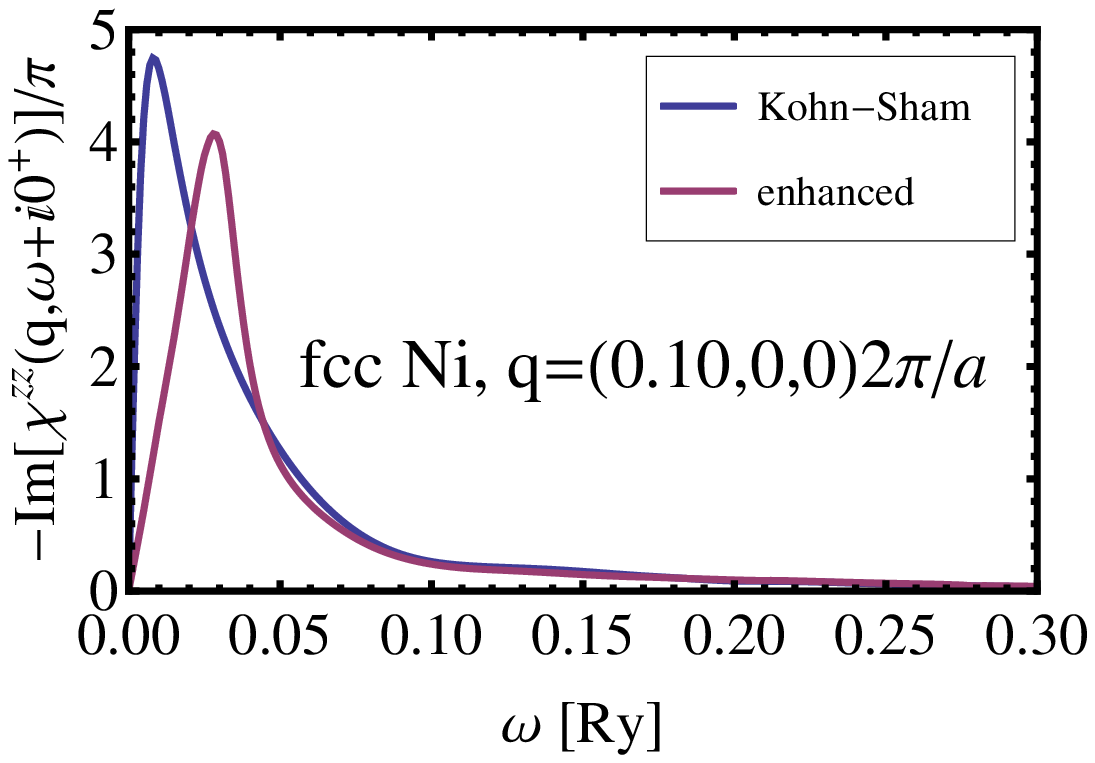}
		\includegraphics[width=0.50\textwidth]{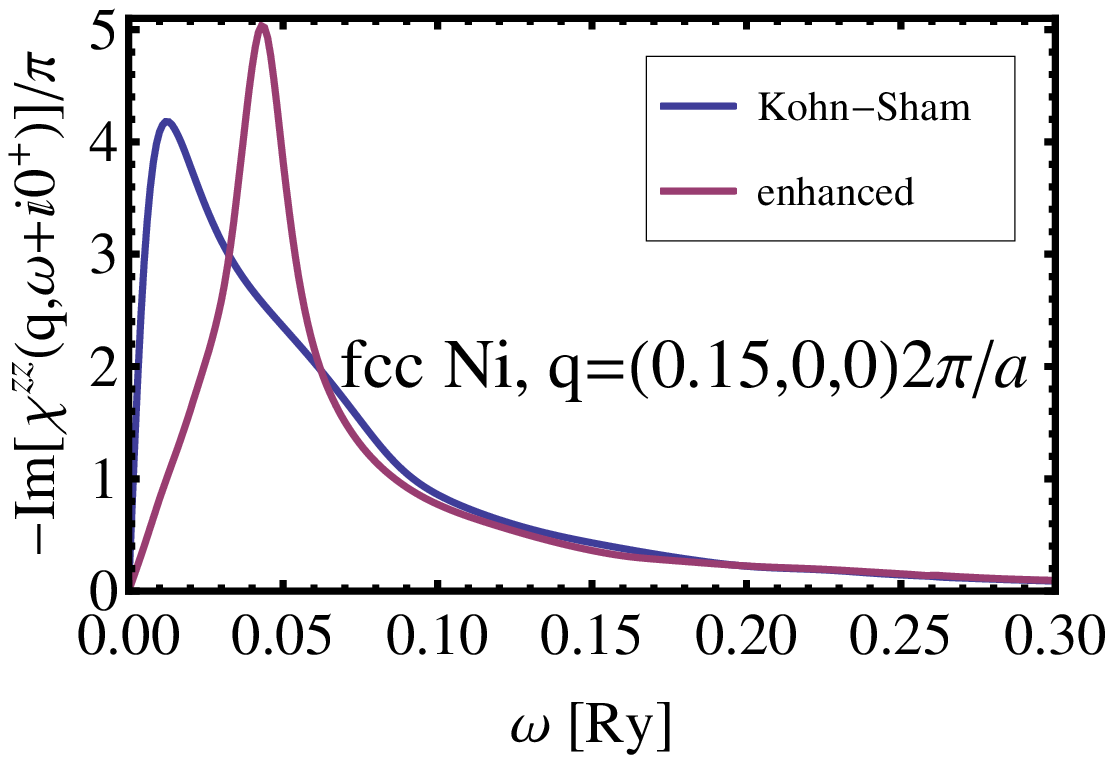}
		\includegraphics[width=0.50\textwidth]{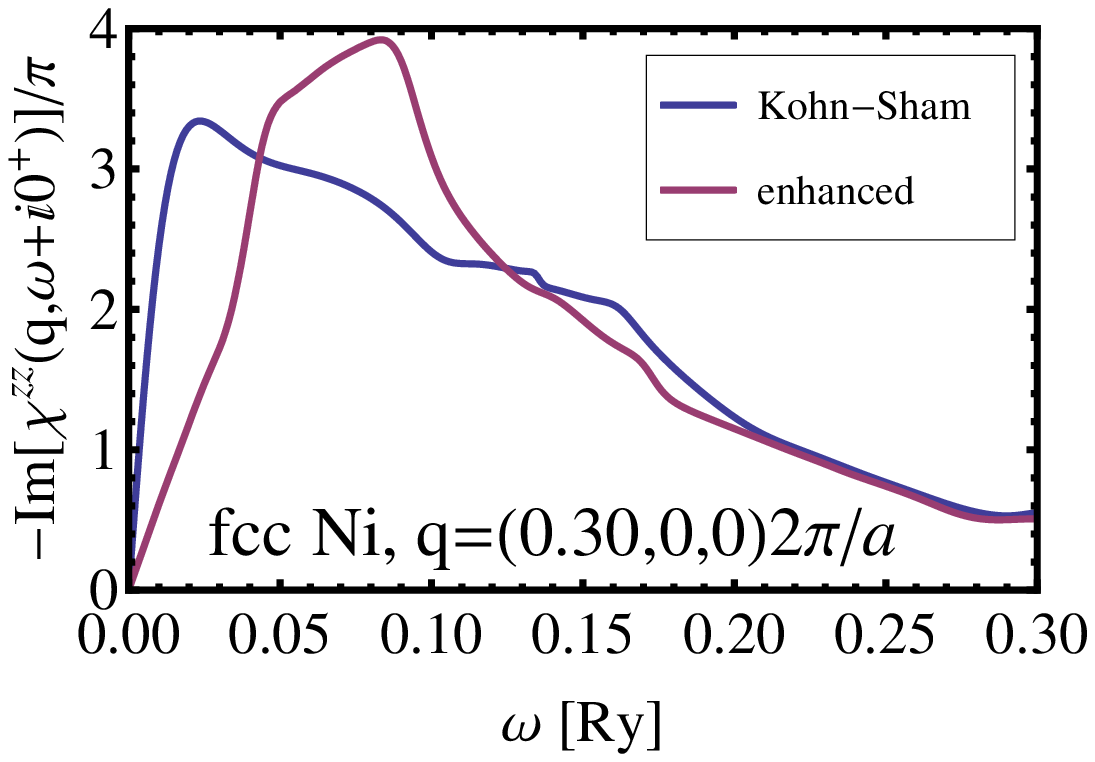}
	\caption{The spin dynamics in the longitudinal channel for fcc Ni and different momenta along the (1,0,0) direction. For simplicity, in the response function, it is assumed that the driving field is uniform in the atomic cell of the atom and the response is integrated over the cell. The dynamics of the formally non-interacting Kohn-Sham system (described by $\suscSymb\KS^{zz}$) is dominated by low energy intraband particle-hole excitations. The dynamics of the fully interacting system, given by the enhanced susceptibility $\suscSymb^{zz}$, features a clear peak for every momentum, appearing somewhat above the Kohn-Sham single particle continuum. The peaks are well defined only for small momenta and loose the well defined above $q \approx 0.3 \times 2\pi/a$.}
	\label{fig:susc-Ni-qx}
\end{figure}

Let us investigate now the details of the spin dynamics in longitudinal channel starting with the fcc Ni presented in Figure~\ref{fig:susc-Ni-qx}. The characteristic pronounced intensity of the single-particle excitations in a narrow low-energy window of width growing with the momentum transfer hints at the dynamics of the formally non-interacting Kohn-Sham system (described by $\suscSymb\KS^{zz}$) dominated by \emph{intraband} particle-hole excitations between hole and electron states of the same band, respectively just below and above the Fermi level \cite{Giuliani2005}. This is in striking contrast to the case of transverse magnetization dynamics of Ni and Fe in which the single particle (Stoner) excitations are formed between states of \emph{different} bands separated by the exchange splitting \cite{Buczek2011a}. The dynamics of the fully interacting system, given by the enhanced susceptibility $\suscSymb^{zz}$, features peaks with well defined energies forming above the Kohn-Sham single particle continuum which, therefore, can be understood as damped, or approximate, eigenstates of the interacting electron liquid. As they form upon the inclusion of electronic interaction, they should be regarded as collective excitations, even though they do not correspond to additional poles of the response function $\suscSymb^{zz}$, contrary to the undamped magnons in the transverse channel. In this respect, they are close cousins of paramagnons in systems like Pd \cite{Doniach1967,Winter1986,Doubble2010} or non-magnetic FeSe \cite{Essenberger2012}. Similar to the case of Pd, the peaks rapidly loose their collective character above $q \approx 0.3 \times 2\pi/a$ and the enhancement of the single particle spectrum becomes marginal. In our understanding, these excitations are the magnetic second sound mode of Ni.

\begin{figure}[htbp]
	\centering
		\includegraphics[width=0.45\textwidth]{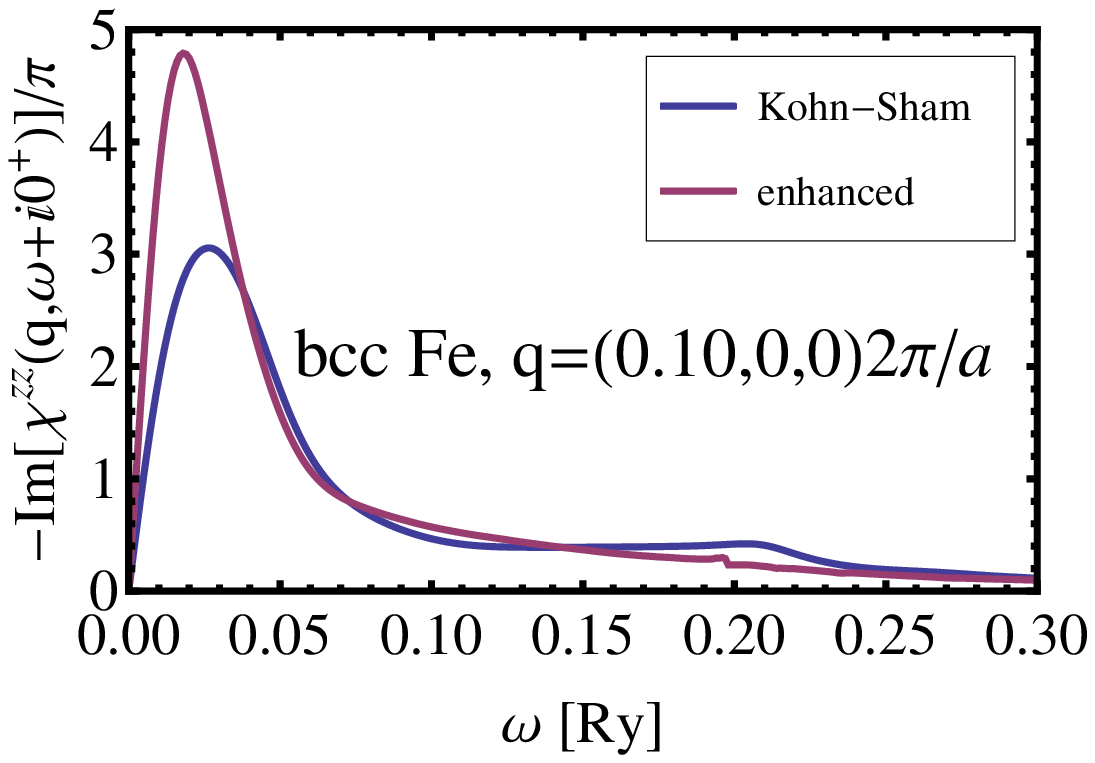}
		\includegraphics[width=0.45\textwidth]{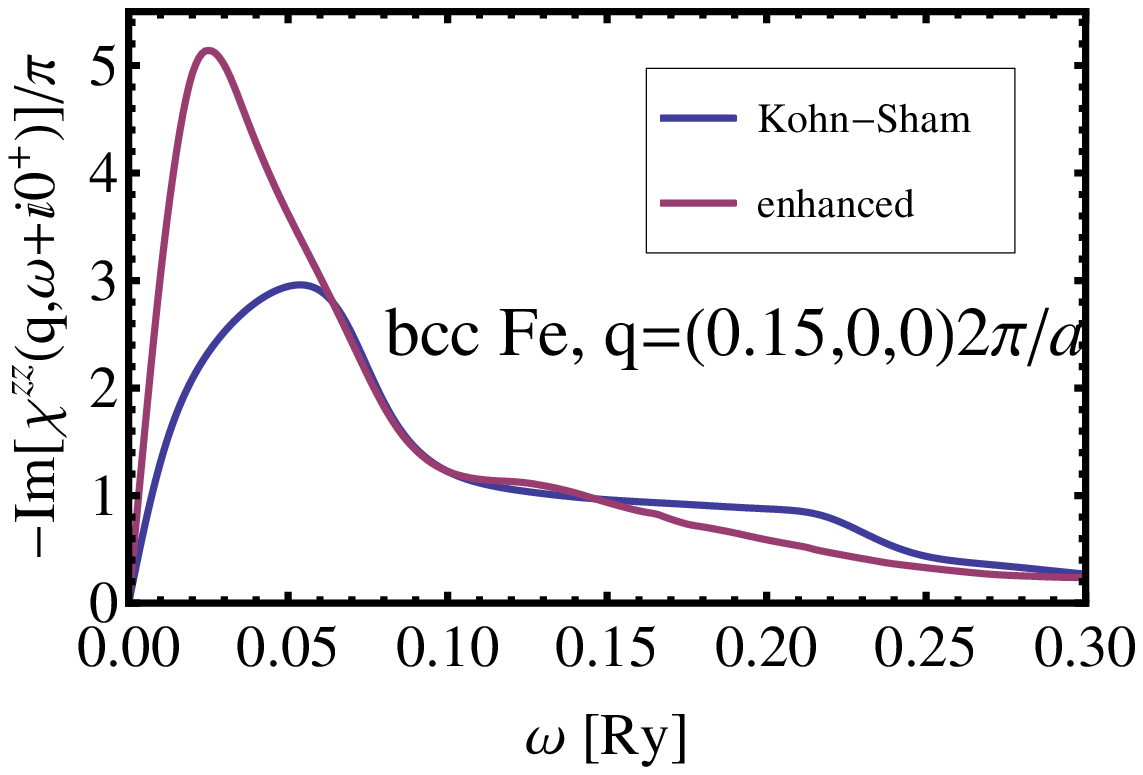}
		\includegraphics[width=0.45\textwidth]{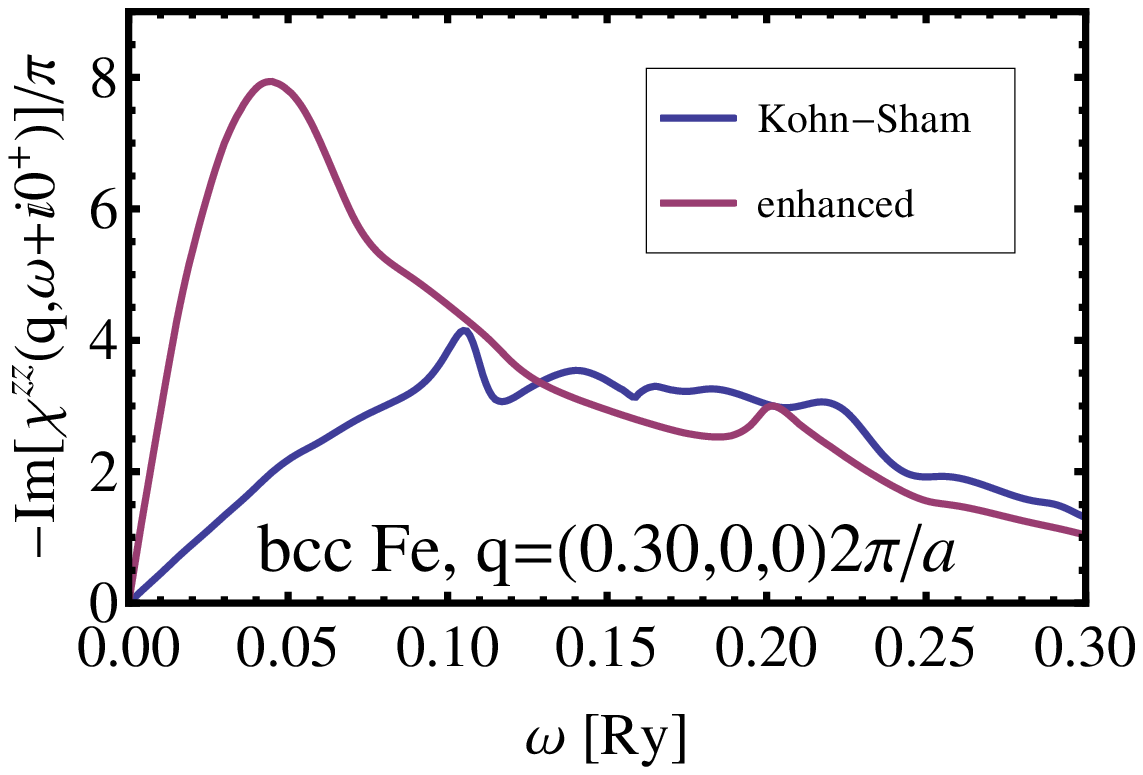}
		\includegraphics[width=0.45\textwidth]{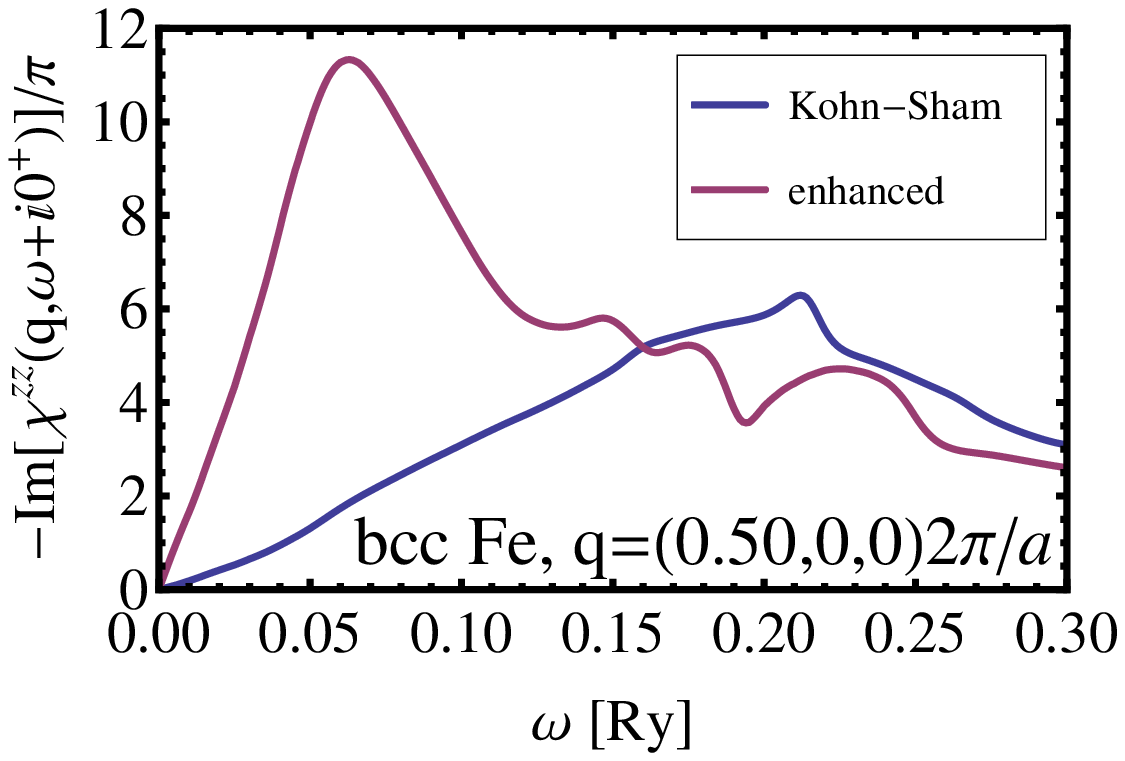}
		\includegraphics[width=0.45\textwidth]{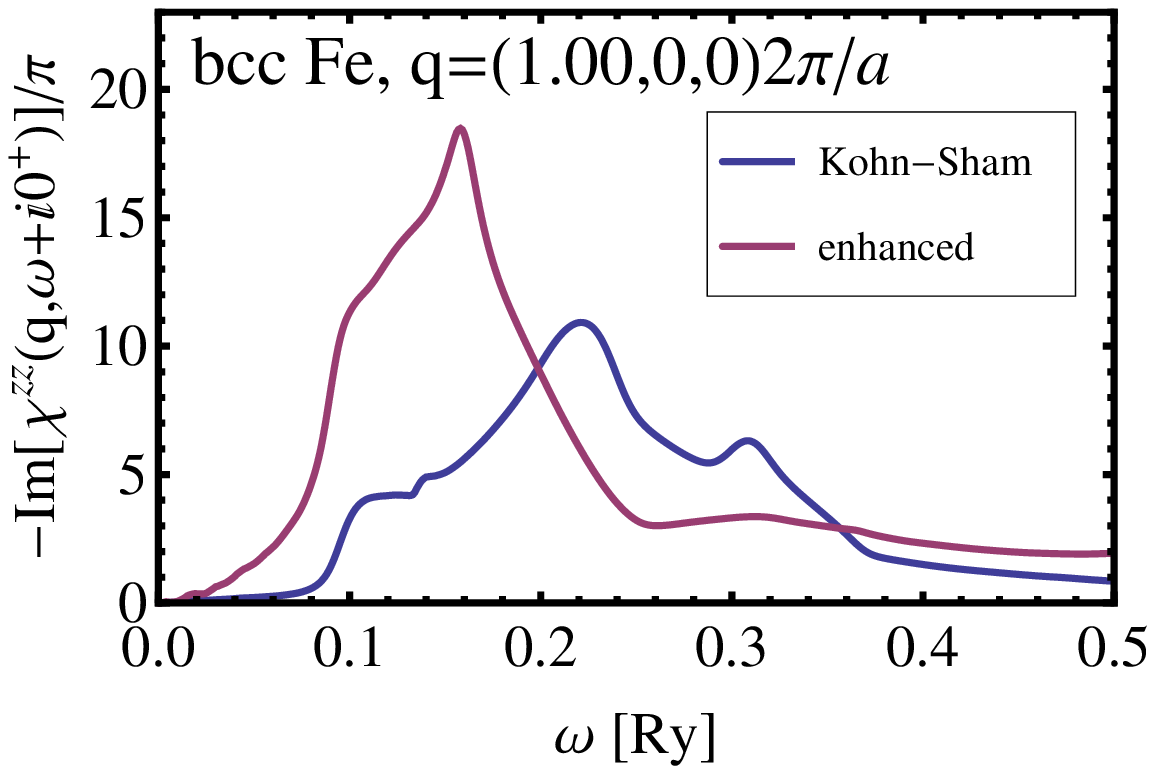}
	\caption{The spin dynamics in the longitudinal channel for bcc Fe and different momenta along the (1,0,0) direction. For simplicity, in the response function, it is assumed that the driving field is uniform in the atomic cell of the atom and the response is integrated over the cell. The dynamics of the formally non-interacting Kohn-Sham system (described by $\suscSymb\KS^{zz}$) is dominated by low energy intraband particle-hole excitations. The dynamics of the fully interacting system, given by the enhanced susceptibility $\suscSymb^{zz}$, features a clear peak for every momentum, appearing somewhat below the Kohn-Sham single particle continuum. The enhancement of the continuum is clearly present in the entire Brillouin zone.}
	\label{fig:susc-Fe-qx}
\end{figure}

Considering to the case of bcc Fe, a very similar picture of the longitudinal spin dynamics emerges, cf.\ Figure~\ref{fig:susc-Fe-qx}. Interestingly, contrary to the case of fcc Ni, the collective mode appears somewhat below the single particle continuum and the enhancement of the continuum is clearly present in the entire Brillouin zone. The origin of this effect cannot be traced back to a single cause. It emerges as an interplay between the shape of the density of the single-particle continuum and the energy dependence of the denominator of the susceptibility Dyson equation which differ in these two systems. This observation stresses the importance of performing the \textit{ab intio} calculations which are able to yield such details without manual fine tuning of model parameters.

It is illustrative to examine the influence of different quantities in the formation of the collective mode in the longitudinal magnetization dynamics channel. Despite of the weakly developed charge excitations in the low energy regime, the coupling terms $\suscSymb\KS^{z0}$ and $\suscSymb\KS^{0z}$ as well as the inclusion of the Coulomb interaction $\vC$ in the susceptibility Dyson equation are indispensable for the proper description of the system dynamics. Considering the $\suscSymb\KS^{zz}$ channel alone, the interaction included in the susceptibility Dyson equation reduces to the exchange-correlation kernel $K\xc$. It this case the denominator of the Dyson equation, $\onemat - \suscSymb\KS^{zz}\fbr{0,0}K\xc$, where $\suscSymb\KS^{zz}\fbr{0,0}$ denotes the static uniform longitudinal susceptibility, features a negative eigenvalue, or, equivalently, an instability in which the system can lower its energy by deforming its charge or magnetization density. Similar effect occurs when the magnetization-charge coupling terms are taken into account but the Coulomb interaction $\vC$ is neglected while keeping $K\xc$ only. (Obviously, neglecting both $\vC$ and $K\xc$ results in $\suscSymb^{zz} = \suscSymb\KS^{zz}$.) This points to the fact that for the corresponding electron system, stripped of the coupling between the charge and magnetization or the Coulomb interaction, the converged ground state is unstable. Obviously, this is unphysical and reflects the crucial role of the Coulomb forces in the stabilization of the magnetic order.

\begin{figure}[htbp]
	\centering
		a)~\includegraphics[width=0.50\textwidth]{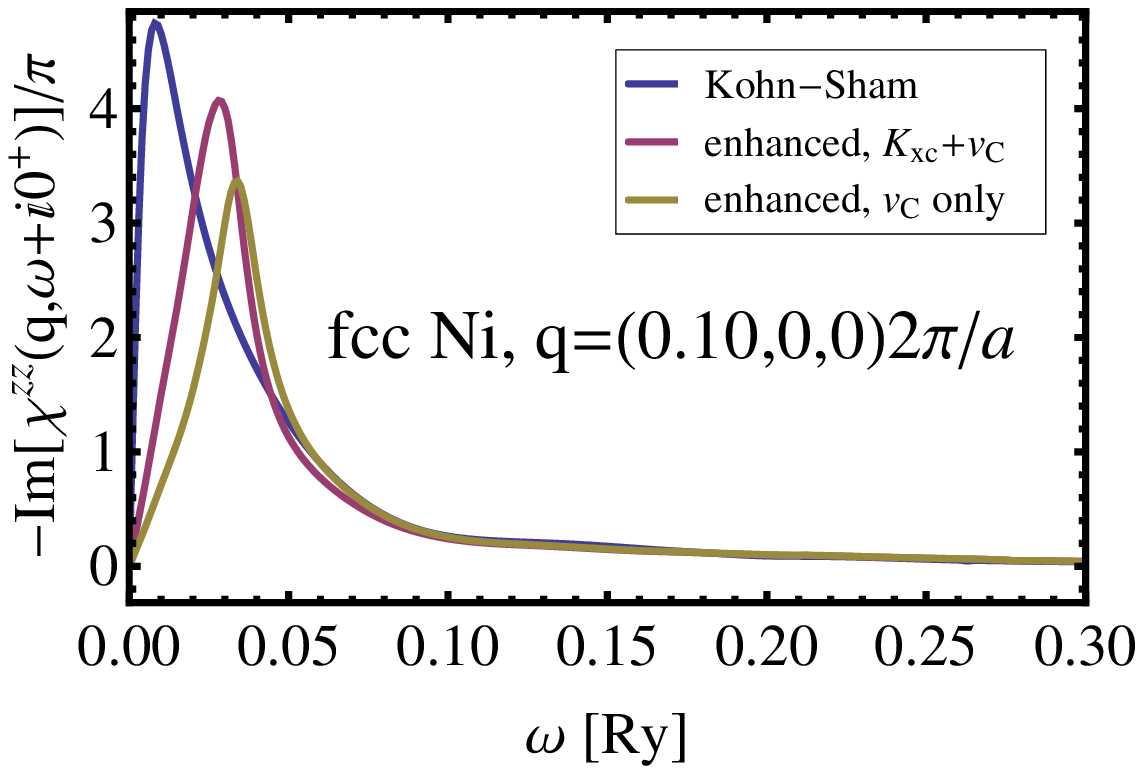}
		b)~\includegraphics[width=0.50\textwidth]{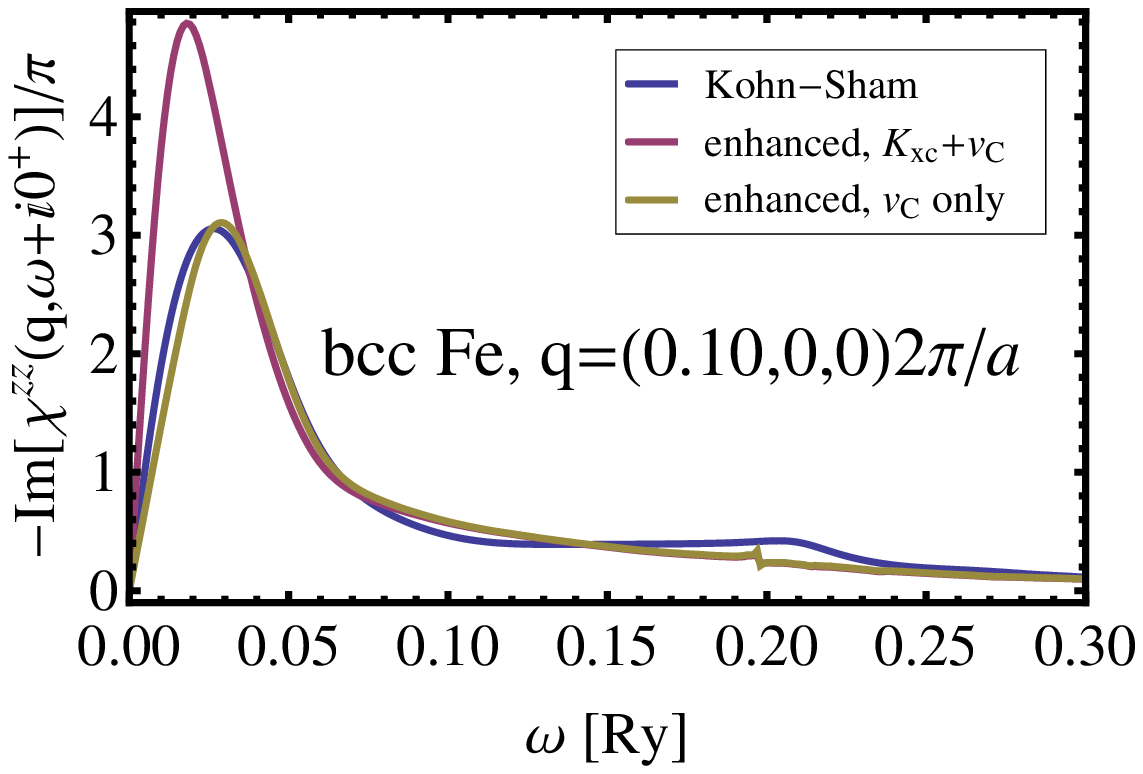}
	\caption{Influence of the exchange correlation kernel $K\xc$ on the collective dynamics in the magnetic longitudinal channel. For simplicity, in the response function, it is assumed that the driving field is uniform in the atomic cell of the atom and the response is integrated over the cell.}
	\label{fig:susc-q0.10-noKxc_comparison}
\end{figure}

On the other hand, the stability of the ground state is preserved when the coupling terms $\suscSymb\KS^{z0}$ and $\suscSymb\KS^{0z}$ are taken into account but the $K\xc$ terms are neglected. In the case of Ni, as evident from Figure~\ref{fig:susc-q0.10-noKxc_comparison}a), this leads to a slight increase of the energy of the collective longitudinal mode which otherwise becomes preserved. The situation is completely different in bcc Fe, cf.\ Figure~\ref{fig:susc-q0.10-noKxc_comparison}b). Here, the $K\xc$ is solely responsible for the enhanced of the Kohn-Sham continuum. This is similar to the case of the transverse magnetization dynamics where the collective modes (magnons) emerge exclusively due to the action of the $K\xc$. We recall briefly that within ALSDA the Coulomb term $\vC$ does not appear in the transverse channel.

\subsection{FeSe}
\label{subsec:FeSe}

In this section, we study the longitudinal magnetic fluctuations in the antiferromagnetic phase of FeSe. Following \cite{Essenberger2012} we use the tetragonal lattice structure with lattice parameters $a = b = 3.765\Angstrom$ and $c = 5.518\Angstrom$. The primitive cell features two Fe atoms residing in the basal plane of the cell and two Se atoms shifted from the plane in the $z$ direction by $\pm \xi c$. In our calculations, we fix $\xi$ to $0.26$. In the local spin density approximation, this setup gives rise to the stable checkerboard antiferromagnetic ground state. In what follows, we denote the sites with magnetization pointing up and down as respectively Fe$_{\uparrow}$ and Fe$_{\downarrow}$.

\begin{figure}[htbp]
	\centering
		\includegraphics[width=0.50\textwidth]{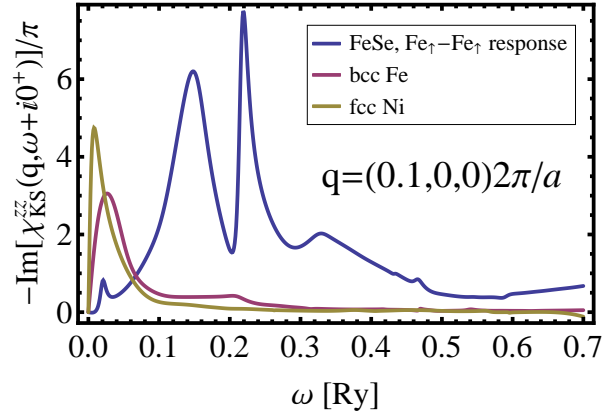}
	\caption{Spectral intensity $-\Im{\suscSymb^{zz}\KS\fbr{\vec{q},\omega + \ii\zp}}/\pi$ of single particle excitations in the formally non-interacting Kohn-Sham system for small momentum transfer $\vec{q} = (0.1,0,0)2\pi/a$. In the case of FeSe system Fe$_{\uparrow}$-Fe$_{\uparrow}$ block of the susceptibility is concerned. For simplicity, in the response function, it is assumed that the driving field is uniform in the atomic cell of the atom and the response is integrated over the cell.}
	\label{fig:chiKS-low_q}
\end{figure}

The longitudinal spin dynamics of the FeSe is very different from the cases of elemental metallic magnets discussed above. Let us first consider the single particle excitations of the formally non-interacting Kohn-Sham system in the $zz$ channel, presented in the Figure~\ref{fig:chiKS-low_q}. For low momentum transfers $\vec{q}$, we observe a series of clear peaks in the energy window between $0.1$\Ry and $0.3$\Ry associated with the \emph{interband} non-spin-flip transitions. The low energy single particle excitations, corresponding below $0.1$\Ry to the \emph{intraband} excitations between states in the close vicinity of the Fermi level, are much weaker. When we examine the corresponding spectra of Fe and Ni in this energy region, we conclude that the characteristic low energy intense intraband excitation signature is practically absent. It is not surprising, taking into account the practically two-dimensional character of the FeSe band structure with its weak dispersion in the $z$ \cite{Essenberger2012} direction and the associated reduced phase space available for the formation of the electron-hole pairs \cite{Giuliani2005}.

\begin{figure}[htbp]
	\centering
		\includegraphics[width=0.50\textwidth]{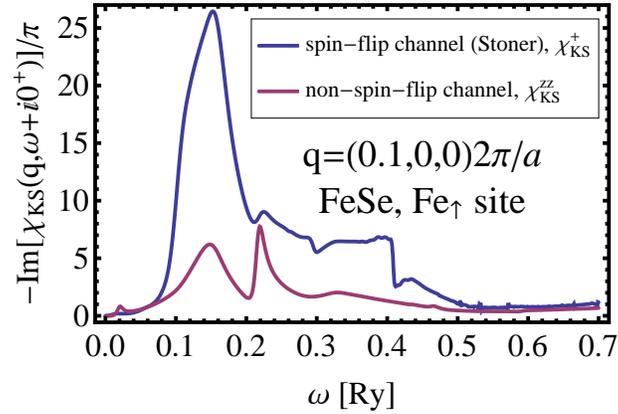}
	\caption{Spectral intensity of single particle excitations in FeSe on the Fe$_{\uparrow}$ site in the formally non-interacting Kohn-Sham system for small momentum transfer $\vec{q} = (0.1,0,0)2\pi/a$. For simplicity, in the response function, it is assumed that the driving field is uniform in the atomic cell of the atom and the response is integrated over the cell. Compared are the spectra of spin-flip (Stoner) and non-spin-flip excitations, given respectively by $\suscSymb^{+}\KS$ and $\suscSymb^{zz}\KS$.}
	\label{fig:chiKS-Stoner_zz_comparison}
\end{figure}

\begin{figure}
	\centering
		a)~\includegraphics[width=0.35\textwidth]{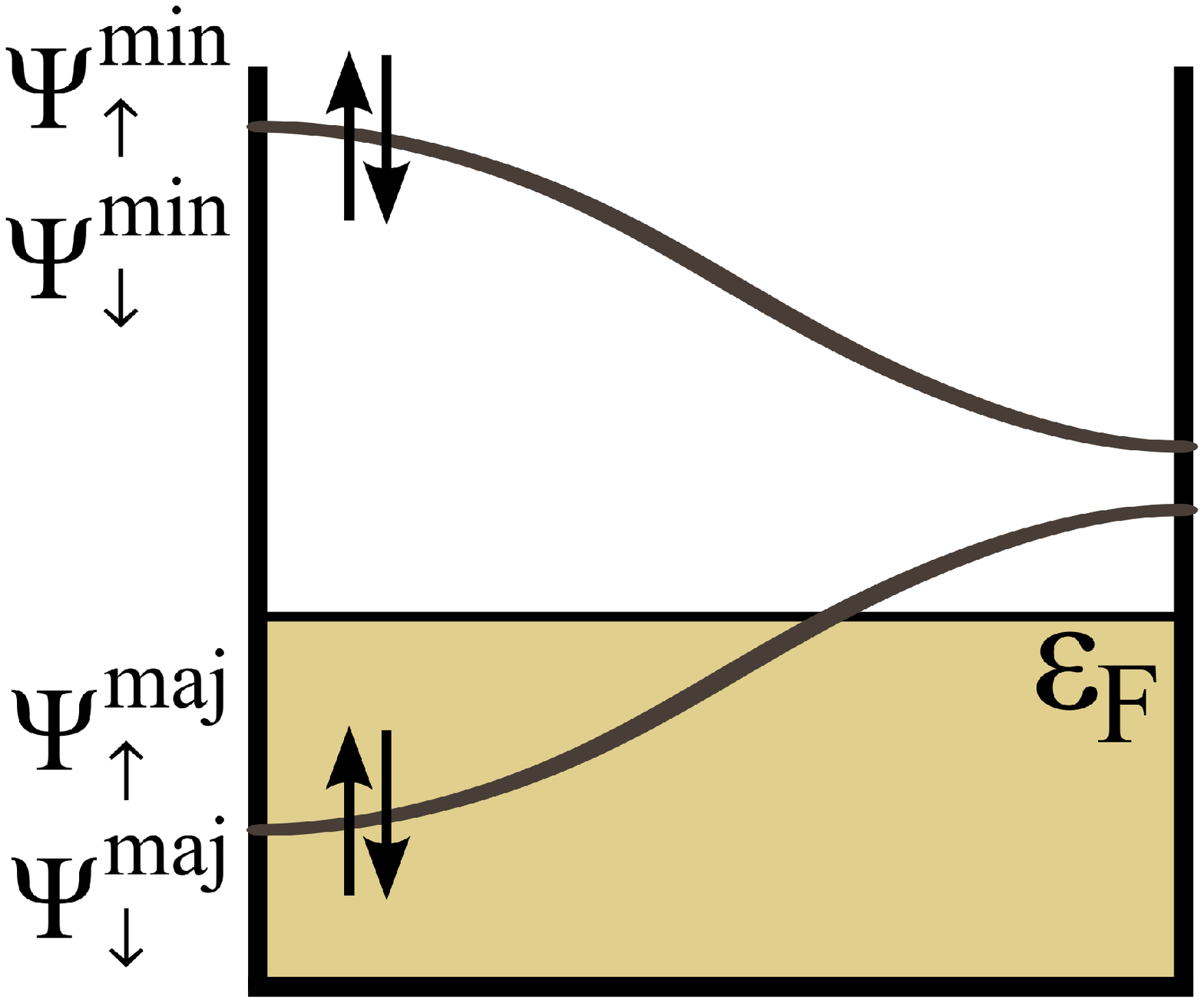}
		b)~\includegraphics[width=0.50\textwidth]{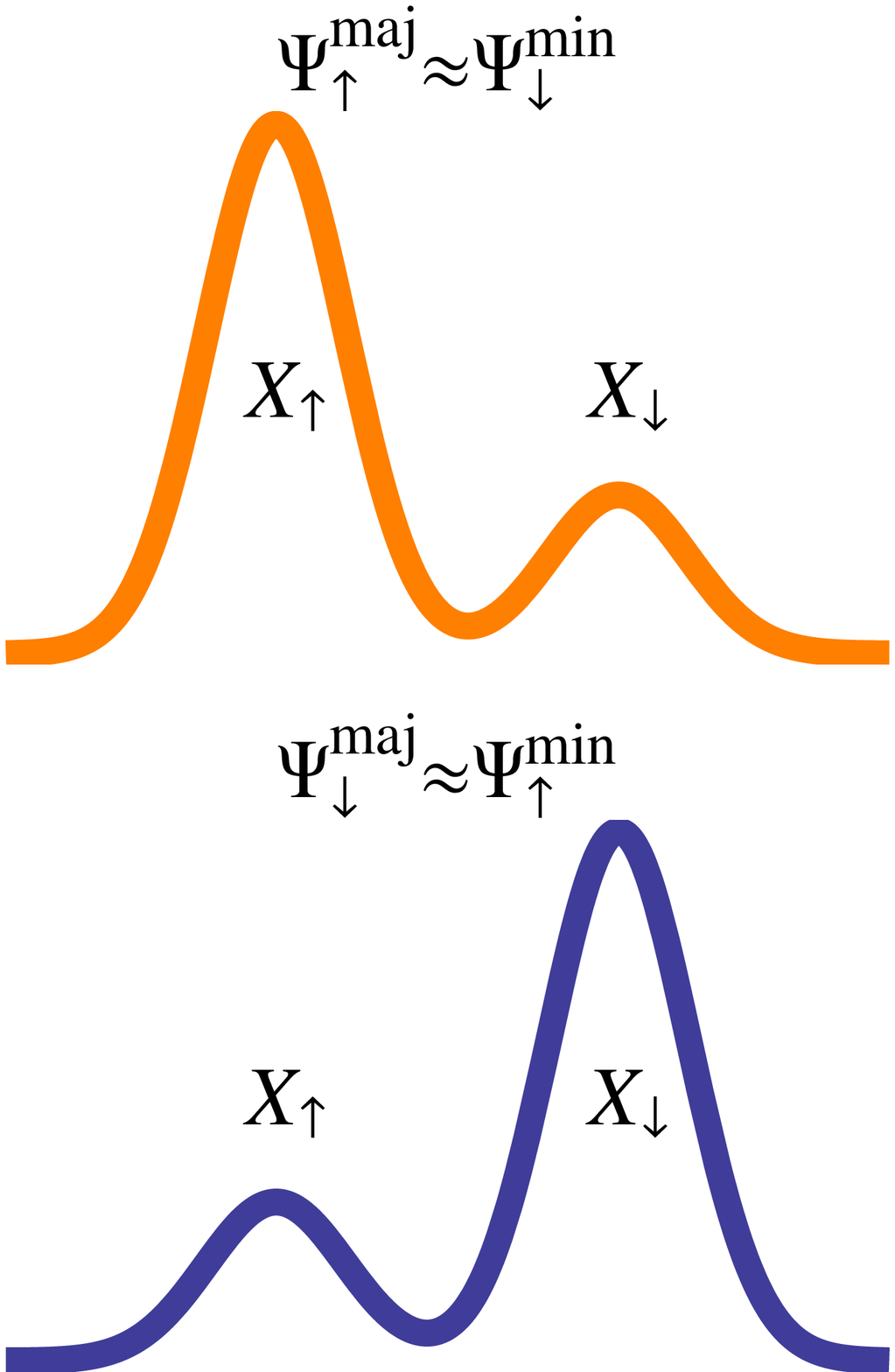}
	\caption{Schematics of bands in an antiferromagnet (a) and their spatial characters (b). $X_{\uparrow}$ and $X_{\downarrow}$ stand for sites with the magnetic moment pointing in the respective direction. Figure (a) is adapted from \cite{Sandratskii2012}.}
	\label{fig:AFM-wave_functions}
\end{figure}

We inquire now which electronic states, respectively below and above the Fermi energy, are involved in the formation of these energetically well defined interband transitions in FeSe. A quick glance at the spectrum of the spin-flip (Stoner) excitations, Figure~\ref{fig:chiKS-Stoner_zz_comparison}, described by the imaginary part of the \emph{transverse} susceptibility $\suscSymb^{+}\KS$, reveals a pronounced intensity appearing in the same energy window. In order to understand this observation, it is necessary to recall the main features of the spin polarized band structure (Kohn-Sham system) of an antiferromagnet \cite{Sandratskii2012}. An antiferromagnet is symmetric with respect to the symmetry transformation consisting of the product of flipping up and down spin states and the space translation transforming one Fe sublattice into other. As a consequence, similar to the case of FeRh or any other antiferromagnet, FeSe features energy degenerate spin-up and -down bands, as shown in the simplified band structure in Figure \ref{fig:AFM-wave_functions}a). Despite this degeneracy, the spatial character of these two types of bands is different, cf.\ the schematics in the Figure \ref{fig:AFM-wave_functions}b). The wave functions of the majority up bands ($\Psi_{\uparrow}^{\textrm{maj}}$) feature higher electron density on the Fe site with the atomic moment pointing up (Fe$_{\uparrow}$) and the wave functions of the majority down bands ($\Psi_{\downarrow}^{\textrm{maj}}$) on the other Fe site with the atomic moment pointing down (Fe$_{\downarrow}$). Additionally, there is a complementary minority spin \emph{down} band $\Psi_{\downarrow}^{\textrm{min}}$ of the spatial character similar to $\Psi_{\uparrow}^{\textrm{maj}}$, i.e. with higher electronic density on the Fe$_{\uparrow}$ site. Because of the presence of the exchange-correlation magnetic field, the state $\Psi_{\downarrow}^{\textrm{min}}$ has higher energy than $\Psi_{\uparrow}^{\textrm{maj}}$ by a value of order of the exchange splitting. The degenerate minority spin up counterpart $\Psi_{\uparrow}^{\textrm{min}}$ feature spatial character similar to $\Psi_{\downarrow}^{\textrm{maj}}$.

\begin{figure}[htbp]
	\centering
		\includegraphics[width=0.50\textwidth]{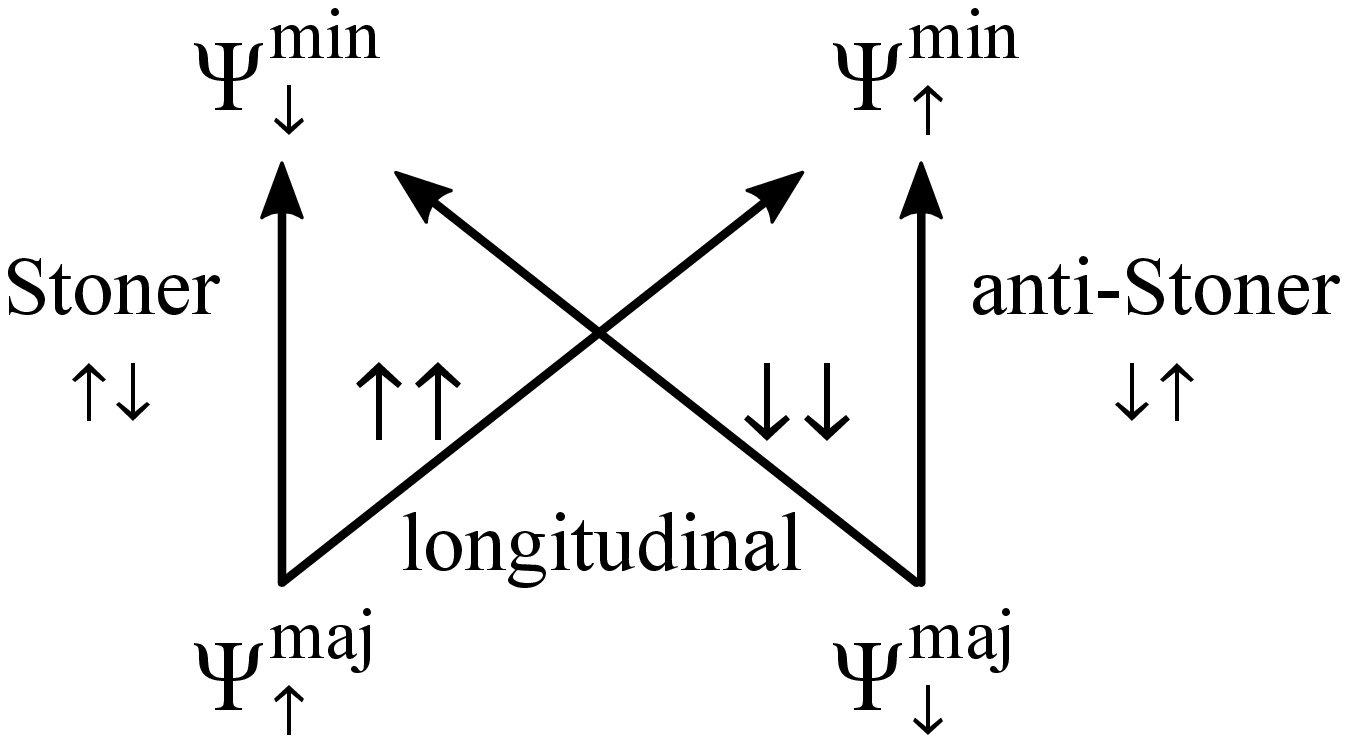}
	\caption{Dominating single particle excitation channels in an antiferromagnet. Stoner and anti-Stoner excitations are described respectively by $\suscSymb^{+}\KS$ and $\suscSymb^{-}\KS$ whereas the longitudinal excitations by $\suscSymb^{\uparrow}\KS$ and $\suscSymb^{\downarrow}\KS$ or equivalently by $\suscSymb^{zz}\KS = \suscSymb^{00}\KS$ and $\suscSymb^{0z}\KS = \suscSymb^{z0}\KS$.}
	\label{fig:AFM-transitions}
\end{figure}

As in the case of ferromagnets, for small momentum transfer, the Stoner excitations in the antiferromagnets are particularly pronounced between spin up and down states separated by the exchange splitting, owing to the similarity of their spatial characters. These transition between $\Psi_{\uparrow}^{\textrm{maj}}$ and $\Psi_{\downarrow}^{\textrm{min}}$ are responsible for the intensive and well defined spectrum of the Stoner excitations in the $0.1$\Ry and $0.3$\Ry energy window in the case of the Fe$_{\uparrow}$ site, cf.\ also Fig.~\ref{fig:AFM-transitions}. The corresponding transitions between energy degenerate counterparts of the wave functions with the same spin, i.e. $\Psi_{\uparrow}^{\textrm{maj}}$ and $\Psi_{\uparrow}^{\textrm{min}}$ as well as $\Psi_{\downarrow}^{\textrm{maj}}$ and $\Psi_{\downarrow}^{\textrm{min}}$, dominate the spectrum of the $\suscSymb^{zz}\KS$. Because of the reduced density associated with $\Psi_{\uparrow,\downarrow}^{\textrm{min}}$ on the Fe$_{\uparrow,\downarrow}$ site, they are necessarily less intense than their Stoner excitations counterparts, as evident from the figure Figure~\ref{fig:chiKS-Stoner_zz_comparison}.

The discussion allows us to conclude that in the case of FeSe the interband transitions between exchange split bands determine the major features of spin dynamics in both the transverse and the longitudinal channel. This is in contrast to the case of elemental 3d ferromagnets, where it is only the case for the transverse channel and the the longitudinal spin dynamics is governed by the intraband low energy non-spin-flip electron excitations.

\begin{figure}[htbp]
	\centering
		\includegraphics[width=0.50\textwidth]{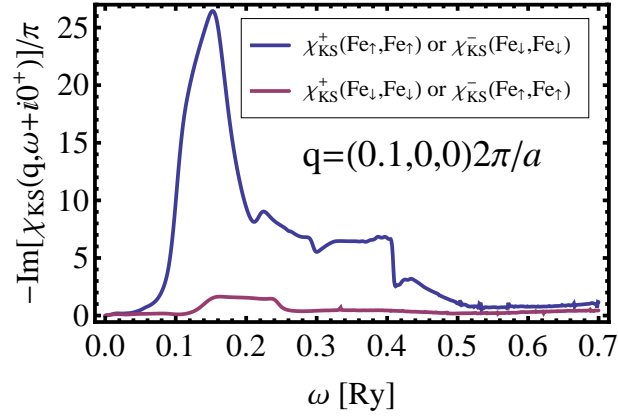}
	\caption{The intensity of Stoner and anti-Stoner excitations in FeSe for low momentum transfer. On the Fe$_{\uparrow}$ the Stoner excitations dominate whereas the Stoner excitations are practically absent. The relation is fully symmetrically opposite on the Fe$_{\downarrow}$ site.}
	\label{fig:Stoner_and_anti-Stoner}
\end{figure}

For the sake of completeness, let us mention that in the transverse channel the antiferromagnets feature both Stoner (spin up to down) and anti-Stoner (spin down to up) excitations appear on an equal footing. The amplitude of the Stoner excitations is pronounced on the sites with magnetization pointing up whereas the anti-Stoner excitations are confined to the sites with magnetization pointing down, as evident from the Figure~\ref{fig:Stoner_and_anti-Stoner}. In the case of the 3d ferromagnets, the anti-Stoner excitation (spin flips from occupied spin minority to empty spin majority bands) are characterized by practically vanishing intensity due to the unavailability of the necessary electron states.

\begin{figure}
	\centering
		\includegraphics[width=0.50\textwidth]{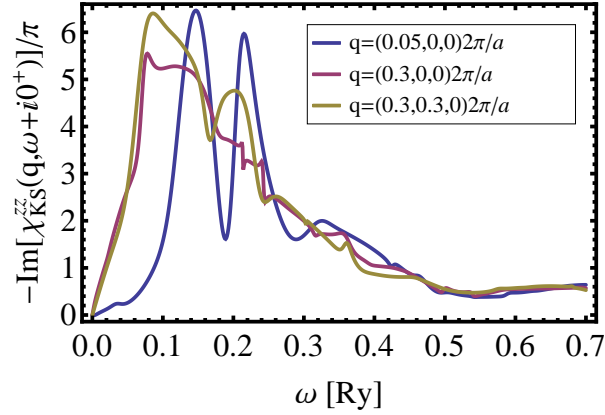}
	\caption{Spectral intensity $-\Im{\suscSymb^{zz}\KS\fbr{\vec{q},\omega + \ii\zp}}/\pi$ of single particle excitations in the formally non-interacting Kohn-Sham system for large momentum transfers compared with the intensity close to the center of the Brillouin zone for FeSe. Fe$_{\uparrow}$-Fe$_{\uparrow}$ block of the susceptibility is concerned. For simplicity, in the response function, it is assumed that the driving field is uniform in the atomic cell of the atom and the response is integrated over the cell.}
	\label{fig:chiKS-large_q}
\end{figure}

With the increasing momentum transfer, the intensity of the single particle excitations in the low energy region increases starting around transfer momentum $q\approx 0.2 \time 2\pi/a$ due to the increase of the available phase space for the formation of electron-hole pairs, in particular for the intraband transitions. For large momenta, cf.\ Figure~\ref{fig:chiKS-large_q}, the well defined peaks vanish and the spectrum is dominated by a wide intensity feature with a maximum around $0.1$\Ry and spreading up to $0.5$\Ry. As we will see later, the qualitative change in the low energy range of the single particle spectra with growing momenta impacts strongly the collective electron dynamics in the longitudinal channel.

\begin{figure}
	\centering
		a)~\includegraphics[width=0.50\textwidth]{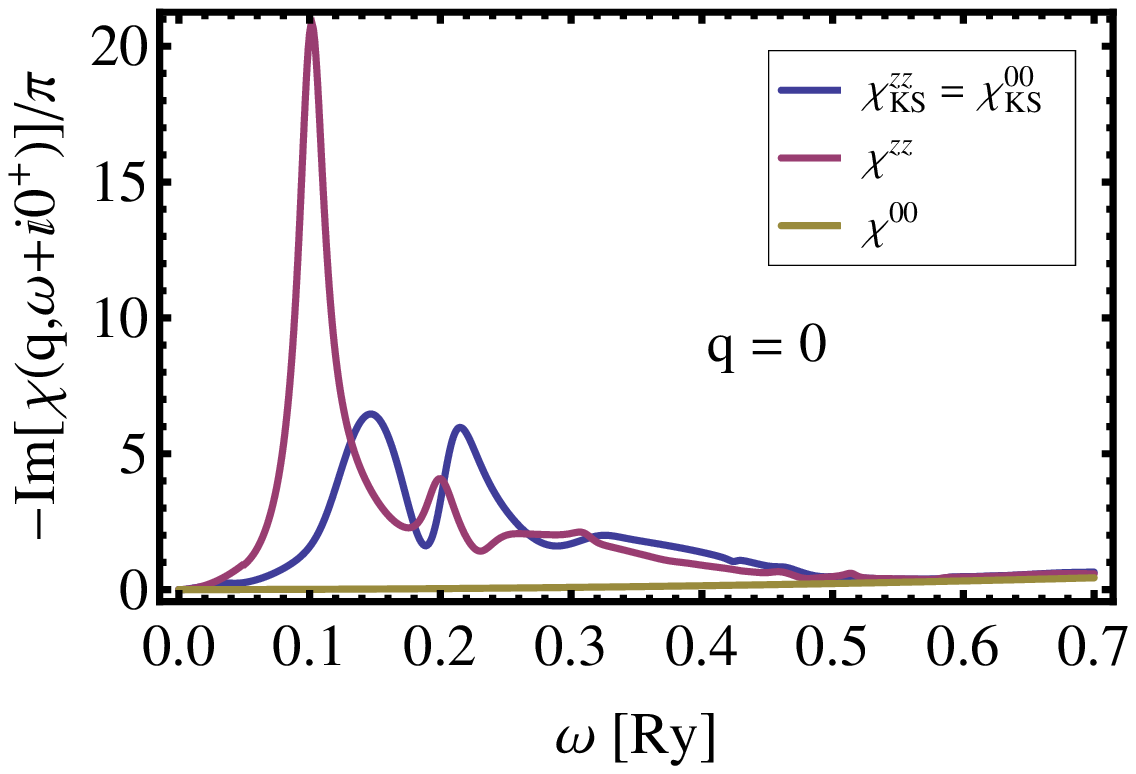}
		b)~\includegraphics[width=0.50\textwidth]{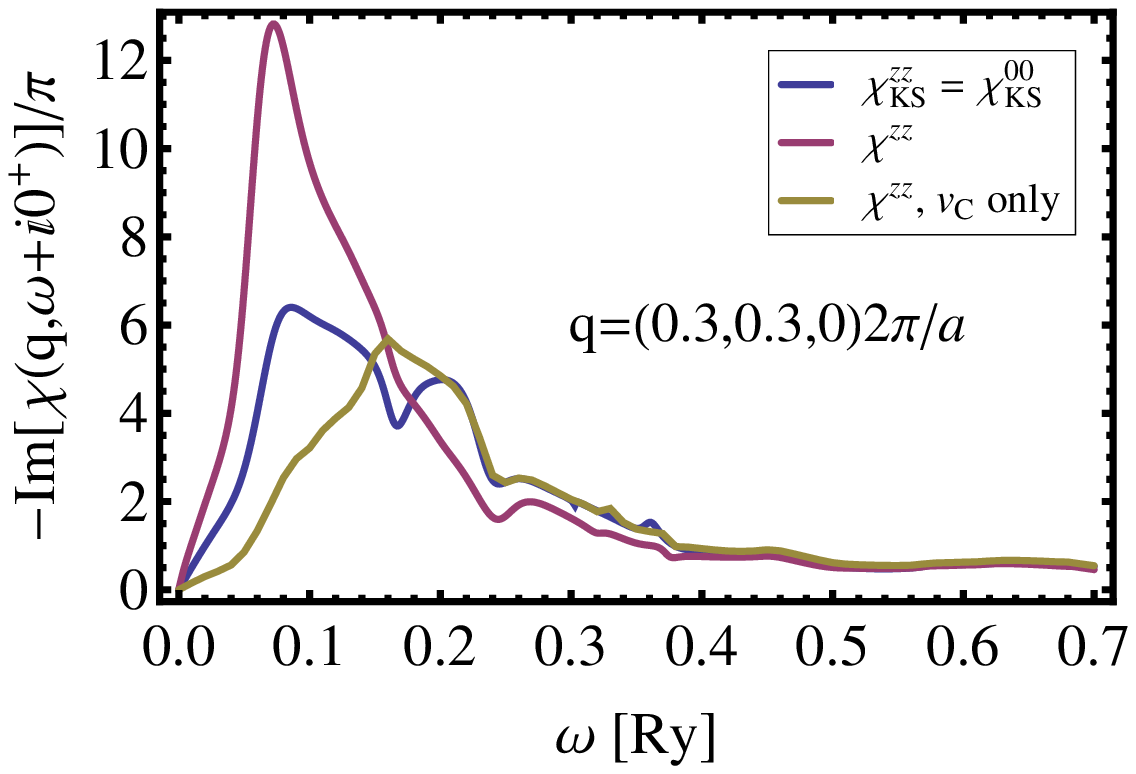}
	\caption{Spectral intensity $-\Im{\suscSymb\fbr{\vec{q},\omega + \ii\zp}}/\pi$ of collective excitations in the longitudinal channel (given by the enhanced susceptibility) compared to the non-interacting Kohn-Sham susceptibility for FeSe. Investigated is the (Fe$_{\downarrow}$,Fe$_{\downarrow}$) or equivalently (Fe$_{\uparrow}$,Fe$_{\uparrow}$) block of the susceptibility for FeSe. In the considered energy range, the clear pronounced second sound peak in the $zz$ channel forms whereas charge-charge susceptibility practically vanishes. Panels (a) and (b) correspond to different momentum transfers. In general, with growing momentum, the energy of the mode decreases slightly and its damping increases quickly above $q\approx 0.25\pi/a$. As example, in panel b), the enhanced susceptibility computed without the $K\xc$ is plotted. Similarly to the case of bcc Fe, the $K\xc$ is indispensable for the enhancement of the Kohn-Sham continuum. Without the inclusion of the kernel, the collective spin dynamics in the longitudinal channel does not form.}
	\label{fig:chiEnh}
\end{figure}

We proceed now to analyze the interacting susceptibility of the FeSe, obtained through the solution of the susceptibility Dyson equation. As in the case of Fe and Ni, in the Kohn-Sham susceptibility the longitudinal $zz$ (spin) and $00$ (charge) channels hybridize strongly ($\suscSymb^{0z}\KS = \suscSymb^{z0}\KS$ is of the same order of magnitude as $\suscSymb^{zz}\KS = \suscSymb^{00}\KS$) but the hybridization is effectively removed at lower energies for the interacting susceptibility, cf.\ Figure~\ref{fig:chiEnh}. In the investigated energy range, the interacting charge susceptibility $\suscSymb^{00}$ is of much smaller magnitude than the longitudinal magnetic susceptibility $\suscSymb^{zz}$. As a consequence, practically free longitudinal magnetic excitations in the $zz$ channel appear whereas the charge dynamics takes place on much higher energy scale and we do not investigate it any further here.

The enhanced susceptibility in the longitudinal $zz$ channel differ strongly from its formally non-interacting Kohn-Sham counterpart. For momenta close to the center of the Brillouin zone, clear pronounced peaks form with maximum around $0.1$\Ry and the full width at half maximum of $0.0244$\Ry, cf.\ Figure~\ref{fig:chiEnh}. This peaks signify strong absorption of energy from the driving monochromatic longitudinal field with this frequency or, alternatively, that the longitudinal magnetic excitation of the electron gas in FeSe is practically an eigenstate of this system, the magnetic second sound. As in the case of Ni and Fe, the nature of this mode is similar to the one of the paramagnon. It is not associated with a new singularity of the $zz$ susceptibility but with the strong \emph{enhancement} of the formally non-interacting particle-hole excitation density. We observe that in the FeSe for low momenta this mode is accompanied by a high energy satellite at around $0.195$\Ry, originating from the structure of the Kohn-Sham intensity.

In the case of FeSe, similarly to the case of bcc Fe, the $K\xc$ is indispensable for the enhancement of the Kohn-Sham continuum. Without the inclusion of the kernel, the collective spin dynamics in the longitudinal channel does not form, as evident from the example Figure~\ref{fig:chiEnh}b).

\begin{figure}[htbp]
	\centering
		\includegraphics[width=0.50\textwidth]{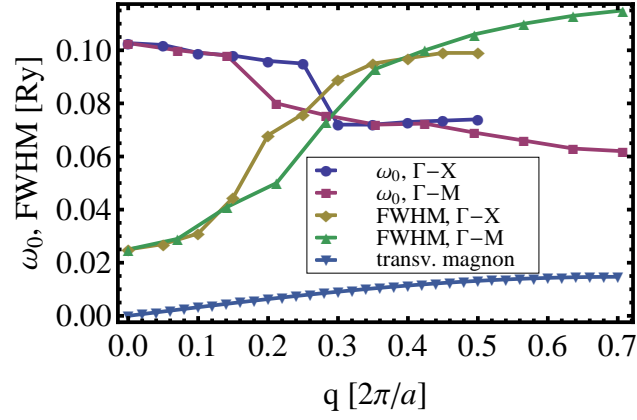}
	\caption{Dispersion of the second sound mode in the FeSe along the two main direction in the Brillouin zone. For different wave vectors, the figure summarizes the position and width of the second sound peaks as exemplarily presented in Figure~\ref{fig:chiEnh}. Clearly visible is the discontinuity around $q\approx 0.25\pi/a$ where the collective mode starts to hybridize strongly with single particle excitations for larger momenta. For comparison, the dispersion of the transverse magnon along the $\Gamma$-M direction \cite{Essenberger2012} is plotted. The lines do not represent data and are only meant as a guide to the eye.}
	\label{fig:FeSe-dispersion}
\end{figure}

As the momentum transfer $\vec{q}$ increases, the energy of the second sound peak shifts gradually to lower energies and its width increases, cf.\ Figure~\ref{fig:FeSe-dispersion}. For $q\approx 0.25\pi/a$, a clear discontinuity in the dispersion of the second sound mode sets in where it enters the region of intense intraband non-spin-flip transitions. This is an example of the Landau damping in which the life-time of a collective mode is strongly reduced by the hybridization with the continuum of single-particle excitations \cite{Buczek2009}. Thus, the momentum dependence of the damping of the second sound mode resembles strongly the spin-wave disappearance mechanism in the transverse channel of materials like bcc Fe \cite{Buczek2011a}.

For comparison, the dispersion of the \textit{transverse} magnon along the $\Gamma$-M direction \cite{Essenberger2012} is included in the Figure~\ref{fig:FeSe-dispersion} as well. We note briefly that the latter theoretical results agree rather nicely with the available inelastic neutron scattering data of Wang \el \cite{Wang2016}. As required by the Goldstone theorem, the dispersion of the transverse magnons starts at $0$ for small momentum transfers, contrary to the longitudinal mode for which the Goldstone boson does not appear. The maximum energy of the transverse magnon, at the edge of the Brillouin zone, amounts to about $0.0147\Ry$. As such, similarly to the case of elemental ferromagnets, the collective longitudinal spin-dynamics develops at substantially higher energies than in the transverse channel.

\begin{figure}[htbp]
	\centering
		\includegraphics[width=0.50\textwidth]{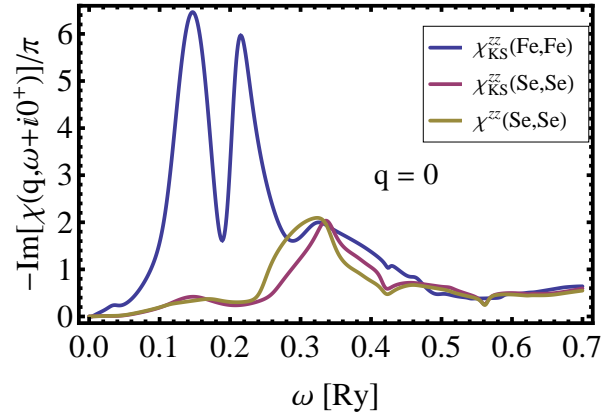}
	\caption{Longitudinal spin dynamics on the Se site. Cf.\ the discussion in the text.}
	\label{fig:FeSe-chiEnh-Se_role}
\end{figure}

Let us shortly discuss the role of the Se in the formation of the second sound mode. Figure~\ref{fig:FeSe-chiEnh-Se_role} presents the intensities of the electron-hole excitations $\suscSymb^{zz}\KS(\textrm{Se},\textrm{Se})$ on the Se site. They are pronounced at much higher energies then their counterparts on the Fe site, $\suscSymb^{zz}\KS(\textrm{Se},\textrm{Se})$. The figure shows that their intensity practically does not change when the Coulomb interaction and $K\xc$ are taken into account. Thus, we conclude that Se is virtually inert in the collective longitudinal spin dynamics.

\section{Discussion and summary}
\label{sec:Summary}

We investigated the longitudinal spin dynamics in elementary 3d ferromagnets bcc and fcc Ni as well as in the antiferromagnetic phase of FeSe. We applied the linear response time dependent density functional theory in order to evaluate the energy and vector dependent charge and spin susceptibility. We resorted to the adiabatic local spin density approximation when evaluating the exchange-correlation kernel.

Contrary to the case of transverse spin excitations (magnons) in collinear magnets, the fluctuations of the longitudinal channel can couple to the charge fluctuations. The coupling is substantial in the formally non-interacting Kohn-Sham case but is effectively lifted by the inclusion of the Coulomb interaction in the interacting susceptibility.

Unlike the transverse magnons, the longitudinal spin fluctuations acquire a collective character without the emergence of a Goldstone boson and are not associated with an additional singularity originating from the susceptibility Dyson equation, similar to the case of paramagnon excitations in non-magnetic metals like Pd. Their properties are strongly material dependent. In Ni and Fe their dynamics is determined by the low energy intraband single particle excitations while in the antiferromagnetic FeSe they derive from the interband transitions within the same family of states which are involved in the formation of the Stoner excitations in the transverse channel. For low momenta, the excitations correspond to well defined peaks in the imaginary part of the susceptibility. In Ni, they gradually loose their collective character with growing momentum, while in Fe they can be observed for momentum transfers in the entire Brillouin zone. The excitations are particularly clearly observable in the antiferromagnetic FeSe close to the center of the Brillouin zone. For growing momenta their life time abruptly shortens around $q\approx 0.25\pi/a$ due to the Landau mechanism, as the excitations enter the region of intense single-particle excitations.

Unfortunately, there is only a limited number of experimental works which we can compare our findings to. The neutron-based experimental investigations of longitudinal spin fluctuations known to us are cited in the introduction. However, they address the energy range well below the window where longitudinal excitations emerge in our calculations. As expected, the experiments do not reveal any well-defined collective modes at these small energies.

There exist several SPEELS studies of inelastic scattering of electrons off different Fe surfaces with energy transfer up to several eV and momentum transfer about halfway to the Brillouin zone boundary \cite{Kirschner1985, Venus1988, Vasilyev2016}. In these experiments, due to the conservation of angular momentum, the signature of the longitudinal magnetization and charge excitations would appear in the non-spin-flip scattering, contrary to the transverse spin-waves which signature is found in the spin-flip channel. To determine this signal, the polarization of outgoing (scattered) electron beam must be analyzed as well which is not the case in most of the recent SPEELS studies \cite{Qin2015}.
For the wave-vectors considered in these experimental studies, we predict the maximum of the magnetic longitudinal mode peak at about 800meV (2eV at the zone boundary, cf.\ Fig.~\ref{fig:susc-Fe-qx}). In older experiments addressing the surface of bcc iron \cite{Kirschner1985, Venus1988} a broad signature of excitations in the non-spin-flip channel is found at about 2eV while in a more recent study of six monolayers thick Fe film on Ir(100) \cite{Vasilyev2016} a clear resonant feature is observed slightly below 5eV. Now, a definite interpretation of these SPEELS experiments is by no means straightforward. Contrary to the case of neutrons, the scattering of electrons is not simply governed by the enhanced response function $\chi$ but may involve a strong signal originating the from single-particle continuum as well ($\chi_{\mathrm{SPEELS}}$) \cite{Vignale1985, Hong2000a} which appear in Fe at somewhat higher energies. Furthermore, our calculations address the bulk Fe while SPEELS is believed to be a surface sensitive method. Thus, the question whether our ALSDA-based theory underestimates the energy of the longitudinal magnetization modes must be deferred to the moment when the dedicated \textit{ab initio} SPEELS cross-section can be computed for these specific surfaces. This is an ongoing effort.

Nevertheless, in Fe, the overall energy scale of the non-spin-flip features found in the SPEELS experiments mentioned would fit the first principles calculations rather well. They appear above the typical energies of spin-waves which in bcc Fe are well defined below 125meV \cite{Buczek2011a}, have energies comparable with the Stoner spin-flip excitation continuum (around 2eV for small momenta) \cite{Sasiouglu2010}, and are order of magnitude less energetic than the charge dynamics in the bulk \cite{Gurtubay2005}.

We drew the analogy between the excitations of the longitudinal magnetization density and the second sound in superfluids based on the observation that both are examples of fluctuating order parameter in a state of spontaneously broken symmetry. However, these two effects feature rather different microscopic details. In the two fluid model of superfluidity, the second sound mode is understood as opposed-phase oscillations of the superfluid and the non-superfluid densities, leaving the total fluid density constant. Contrary to the latter effect, in the ``first sound'' mode, associated with the excitation of the ``usual'' phonons, the two densities oscillate in phase, changing locally the fluid density but keeping the ratio of superfluid and the non-superfluid densities constant. In a magnet, the analog of the phonons are the transverse magnons.

In an itinerant ferromagnet, at absolute zero, the role of the superfluid density, i.e. the order parameter, is played by the magnetization density, $m^{z} = n^{\uparrow} - n^{\downarrow}$, whereas the normal density, i.e. the unpolarized fraction of the electron liquid, is given by $\rho^{0} = 2 n^{\downarrow}$. Obviously, the two densities add up to the total local electron density $n^{0} = n^{\uparrow} + n^{\downarrow}$. (In the antiferromagnetic case, the definition of the up and down reference directions must be allowed to vary in space following the ground state magnetization direction.) As shown by our preceding analysis of itinerant magnets, in the longitudinal mode the $m_{z}$ and $\rho_{0}$ do not necessarily oscillate with opposite phases. This is particularly the case when the non-interacting (KS) susceptibility is concerned. In this case, the fluctuations described by $\suscSymb^{\uparrow}$ and $\suscSymb^{\downarrow}$ are independent from each other giving rise to two distinct longitudinal normal modes, both involving strongly coupled dynamics of spin and charge densities. However, these up and down density channels decouple only in the formally non-interacting Kohn-Sham single particle picture. In the time depend density functional framework, the density induced in an excitation of the electron gas alters the effective Kohn-Sham potential. This effect connects the dynamics of the two spin channels and, when the interacting susceptibility is concerned, the picture of the longitudinal modes changes drastically. The Coulomb interaction is responsible for the formation of plasmons in the charge density channel with typical energies well above those of the longitudinal magnetic fluctuations. \footnote{We remark in passing that the plasmons are Goldstone bosons arising due to the spontaneously broken symmetry of Galilei boosts with their mass originating from the longrangeness of the Coulomb interaction \cite{Morchio1986}.} For low frequencies, the true (interacting) magnetization susceptibility ($\suscSymb^{zz}$) clearly dominates the other terms in the response function ($\suscSymb^{00}$, $\suscSymb^{0z}$, and $\suscSymb^{z0}$). This separation of the energy scales practically decouples the longitudinal magnetization and charge dynamics. Effectively, similar to the case of the liquid helium, the order parameter in magnets tends to oscillate without affecting the total particle density.

We briefly add that this picture in magnets changes again at elevated temperatures below the phase transition. The normal fluid density increases by the density of thermally excited magnons which determines the magnitude of the reduction of the $z$-component of the magnetization relative to the magnetization at $T = 0$. In this case the magnetic second sound involves, in addition to the electronic density fluctuations present even at $T = 0$, a sound wave within the gas of magnon excitations. Contrary to the case of the spin and electron density dynamics at absolute zero, the latter effect can be captured already within the Heisenberg Hamiltonian models with rigid magnetic moments. This effect is a subject of a separate study.

We believe that longitudinal spin fluctuations may come to play a prominent role in the field of spintronics \cite{Chumak2015}, i.e. the computer engineering area using magnetic degrees to freedom to perform logical operations. One of the key challenges in this field is the construction of "transducers", i.e. devices allowing different degrees of freedom to effectively couple to the spin dynamics and thus allowing us to exercise a reliable control over it. Using charge current, or electric field, would be attractive when interfacing the magnonic logic gates with conventional computers. In collinear magnets, in linear regime, spin-flip and charge excitations can couple practically only via the relatively weak spin-orbit interaction. On the other hand, the spin excitations in the longitudinal channel would strongly interact with plasmons in the systems in which the two classes of excitations appeared in the same energy window. \textit{Ab initio} studies can pave the way to engineer such materials. Spin-flip magnetic excitations have been recently shown to couple to phonons \cite{Weiler2011, Thevenard2014, Berk2019} offering an alternative way to interact with the spin dynamics. It seems conceivable that the spin dynamics in the longitudinal channel could couple even stronger to the phonons than its spin-flip counterpart owing to its character hybridized with charge excitations. However, the coupling would be most probably possible only for optical phonons due to the comparably high energies of the longitudinal spin fluctuations.

We hope that our investigations into the nature of longitudinal spin excitations will stimulate further theoretical and experimental studies in this vastly unexplored domain.

\section{Acknowledgments}
\label{sec:Acknowledgments}

P.B. acknowledges the most generous support of the Alexander von Humboldt Foundation and the hospitality of the University of Missouri-Columbia, Columbia, Missouri, USA, and Max Planck Institute for Microstructure Physics, Halle (Saale), Germany, during the conduction of this research in spring 2012.

\bibliographystyle{apsrev}
\bibliography{../bibliography/magnetism}

\end{document}